\documentclass[%
 showpacs,
 amsmath,amssymb,
]{revtex4-1}

\usepackage{graphicx}
\usepackage{dcolumn}
\usepackage{bm}

\usepackage{CJK}

\usepackage{color}
\usepackage{amssymb}
\usepackage{amsfonts}

\usepackage[colorlinks,urlcolor=blue,linkcolor=blue,citecolor=blue,anchorcolor=blue]{hyperref}

\begin{document}

\preprint{APS/123-QED}

\title{Edge states in dynamical superlattices}

\author{Yiqi Zhang$^1$}
\email{zhangyiqi@mail.xjtu.edu.cn}
\author{Yaroslav V. Kartashov$^{3,4,5}$}
\author{Feng Li$^{1}$}
\author{Zhaoyang Zhang$^{1,2}$}
\author{Yanpeng Zhang$^{1}$}
\email{ypzhang@mail.xjtu.edu.cn}
\author{Milivoj R. Beli\'c$^{6}$}
\author{Min Xiao$^{7,8}$}
\affiliation{%
 $^1$Key Laboratory for Physical Electronics and Devices of the Ministry of Education \& Shaanxi Key Lab of Information Photonic Technique,
Xi'an Jiaotong University, Xi'an 710049, China \\
$^2$Department of Applied Physics, School of Science, Xi'an Jiaotong University, Xi'an 710049, China\\
$^3$ICFO-Institut de Ciencies Fotoniques, The Barcelona Institute of Science and Technology, 08860 Castelldefels (Barcelona), Spain\\
$^4$Institute of Spectroscopy, Russian Academy of Sciences, Troitsk, Moscow Region 142190, Russia\\
$^5$Department of Physics, University of Bath, Bath BA2 7AY, United Kingdom\\
$^6$Science Program, Texas A\&M University at Qatar, P.O. Box 23874 Doha, Qatar \\
$^7$Department of Physics, University of Arkansas, Fayetteville, Arkansas, 72701, USA\\
$^8$National Laboratory of Solid State Microstructures and School of Physics, Nanjing University, Nanjing 210093, China
}%

\date{\today}

\begin{abstract}
\noindent
  We address edge states and rich localization regimes available in
the one-dimensional (1D) dynamically modulated superlattices, both
theoretically and numerically. In contrast to conventional lattices
with straight waveguides, the quasi-energy band of infinite modulated
superlattice is periodic not only in the transverse Bloch momentum,
but it also changes periodically with increase of the coupling strength
between waveguides. Due to collapse of quasi-energy bands dynamical
superlattices admit known dynamical localization effect. If, however,
such a lattice is truncated, periodic longitudinal modulation leads
to appearance of specific edge states that exist within certain periodically
spaced intervals of coupling constants. We discuss unusual transport
properties of such truncated superlattices and illustrate different
excitation regimes and enhanced robustness of edge states in them,
that is associated with topology of the quasi-energy band.
\end{abstract}

\maketitle

\section{Introduction}

Periodically modulated lattice systems attract considerable attention
in diverse areas of physics, including condensed matter physics \cite{oka.prb.79.081406.2009,kitagawa.prb.82.235114.2010,lindner.np.7.490.2011,rudner.prx.3.031005.2013,leon.prl.110.200403.2013,goldman.prx.4.031027.2014,asboth.pra.91.022324.2015,xiong.prb.93.184306.2016,titum.prx.6.021013.2016}
and photonics \cite{yu.np.3.91.2009,fang.np.6.782.2012,kitagawa.nc.3.882.2012,khanikaev.nm.12.233.2012,liang.prl.110.203904.2013,rechtsman.nature.496.196.2013,chen.nc.5.5782.2014,leykam.prl.117.013902.2016,leykam.prl.117.143901.2016,weimann.nm.16.1476.2017}.
One of the main reasons behing interest to such systems is that due
to variation of parameters of the system along the evolution coordinate
(time in condensed matter physis or propagation distance in photonics)
not only rich variety of resonant dynamical effects associated with
specific deformations of quasi-energy bands appears (for an overview
of such dynamical effects see \cite{longhi.lpr.3.243.2009,garanovich.pr.518.1.2012}), but one
may also encounter the effects of purely topological origin. One of
the manifestations of such effects is the appearance of topologically
protected edge states that are typically unidirectional (in the 2D
systems) and that demonstrate immunity to backscattering on disorder
and other structural lattice defects due to topological protection.
In modulated periodic photonic systems, frequently called Floquet
insulators \cite{lindner.np.7.490.2011,rechtsman.nature.496.196.2013,zhang.lpr.9.331.2015,maczewsky.nc.8.13756.2017},
longitudinal variations of underlying potential
were shown to lead to appearance of the effective external time-dependent
``magnetic fields'' that qualitatively change the behaviour of the
system and allow to design a new class of devices employing topologically
protected transport, including photonic interconnects, delay lines,
isolators, couplers, and other structures \cite{maczewsky.nc.8.13756.2017}.
Periodically modulated photonic lattices were employed for realization
of discrete quantum walks \cite{broome.prl.104.153602.2010,sansoni.prl.108.010502.2012},
and allowed observation of Floquet topological transitions with matter
waves \cite{asboth.prb.90.125143.2014,lago.pra.92.023624.2015}.

Previous investigations of modulated lattices were mainly focused
on the 2D and 3D geometries, and less attention was paid to the 1D
settings. Moreover, upon consideration of bulk and surface effects
in the modulated photonic 1D systems only simplest lattices were utilized
with identical coupling strength between all channels and with identical
(usually sinusoidal) laws of their longitudinal variation \cite{longhi.prl.96.243901.2006,garanovich.prl.100.203904.2008,szameit.prl.101.203902.2008,longhi.oc.281.4343.2008,szameit.prl.102.153901.2009,longhi.prb.80.235102.2009,kartashov.ol.34.2906.2009}. Only recently dynamical \textit{superlattices}
with specially designed periodically varying separation between channels
belonging to two different sublattices were introduced that allowed
observation of intriguing new resonant phenomena, such as light rectification \cite{longhi.ol.34.458.2009,dreisow.epl.101.44002.2013,kartashov.pra.93.013841.2016}.
Previously only bulk modulated superlattices were considered
and no surface effects in such structures were addressed. Therefore,
main aim of this work is the exploration of new phenomena stemming
from the interplay between superlattice truncation and its longitudinal
modulation. We aim to show that dynamically modulated \textit{truncated}
superlattices exhibit topological transition manifested
in qualiative modification of the quasi-energy spectrum upon variation
of the coupling strength between waveguides forming the lattice. Namely,
within proper intervals of coupling strength the isolated eigenvalues
appear that are associated with nonresonant (i.e. existing within
continuous intervals of coupling strengths) edge states. Interestingly,
such edge states persist even when conditions for collapse of bulk
quasi-energy band are met. We discuss specific propagation dynamics
in the regime, where edge states exist. We believe that these findings
substantially enrich the approaches for control of propagation paths
of light beams in periodic media.

\begin{figure}[htpb]
\centering \includegraphics[width=0.5\columnwidth]{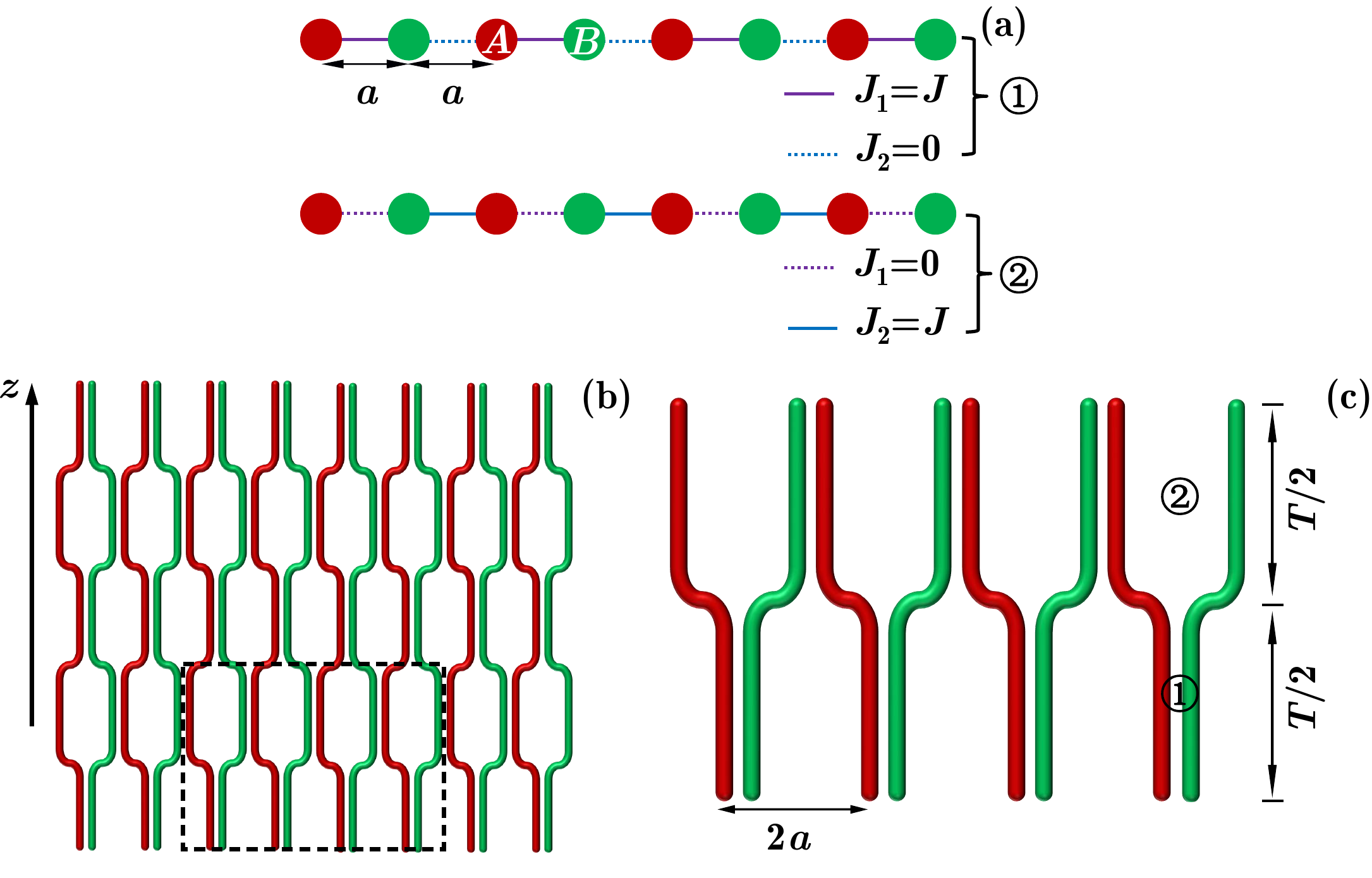}
\caption{(a) Shematic illustration showing coupling constants on two half-periods
of superlattice composed from sublattices A and B. Coupling only occurs
between sites connected by solid lines and is absent between sites
connected by dashed lines. (b) Refractive index distribution in photonic
lattice that reproduces coupling scheme illustrated in panel (a).
(c) Magnification of the region marked by a dashed box in (b) that
shows one longitudinal period of the structure.}
\label{fig1}
\end{figure}

As an example of the dynamical superlattice we consider discrete structure
depicted in Fig. \ref{fig1}, which is somewhat similar to the Su-Schrieffer-Heeger
lattice \cite{asboth.book.2016}. The superlattice is composed of
two sublattices, denoted as $A$ and $B$ (red and green channels
in Fig. \ref{fig1}). The single-mode waveguides in individual sublattices are
curved such that coupling strength between nearest neighbours belonging
to different sublattices changes with propagation distance in a step-like
fashion, as shematically shown in Fig. \ref{fig1}(a) {[}since there are two
sublattices, one can introduce two coupling strengths $J_{1}(z)$
and $J_{2}(z)$ describing coupling between waveguides with equal
$(n,n)$ or with different $(n,n+1)$ indices from two sublattices{]}.
We assume that the coupling strength increases to maximal value $J$
when two waveguides are close and drops down nearly to zero when they
are well separated, due to exponential decrease of the overlap integrals
between modal fields with increase of the distance between waveguides.
The longitudinal period of the structure is given by $T$, while transverse
period is given by $2a$. In Fig. \ref{fig1}(c) we display one longitudinal
period of the structure indicated by a dashed box in Fig. \ref{fig1}(b).
Coupling constants on two different segments of the lattice are indicated
in Fig. \ref{fig1}(a). Such a lattice can be easily fabricated with
femtosecond-laser writing technique \cite{dreisow.epl.101.44002.2013,rechtsman.nature.496.196.2013,plotnik.nm.13.57.2014,vicencio.prl.114.245503.2015,mukherjee.prl.114.245504.2015,diebel.prl.116.183902.2016}.

\section{Theoretical model and band structure}

We describe propagation of light in the infinite superlattice depicted
in Fig. \ref{fig1} using discrete model \cite{szameit.pra.84.021806.2011,weimann.ol.41.2414.2016}
\begin{equation}
\begin{split}i\frac{dA_{n}}{dz}= & J_{1}(z)B_{n}+J_{2}(z)B_{n-1},\\
i\frac{dB_{n}}{dz}= & J_{1}(z)A_{n}+J_{2}(z)A_{n+1},
\end{split}
\label{eq1}
\end{equation}
where coupling constants $J_{1,2}(z)$ are step-like periodic functions
of the propagation distance $z$, while $A_{n},\,B_{n}$ stand for the
field amplitudes on sites of sublattices $A$ and $B$. According to the
Floquet theory, the evolution of excitations in longitudinally modulated
lattice governed by the Hamiltonian $H(\mathbf{k},t)=H(\mathbf{k},t+T)$
(here $T$ is the period of longitudinal modulation and $\mathbf{k}$
is the transverse Bloch momentum), can be described by the Floquet
evolution operator $U(t)=\mathcal{T}\exp[-i\int_{0}^{t}H({\bf k},t')dt'],$
where $\mathcal{T}$ is the time-ordering operator. Defining evolution
operator $U(T)$ for one longitudinal period of the structure {[}i.e.
$|\phi(\mathbf{k},T)\rangle=U(T)|\phi(\mathbf{k},0)\rangle$, where
$|\phi(\mathbf{k},t)\rangle$ is the Floquet eigenstate of the system{]}
and using adiabatic approximation, one can introduce effective Hamiltonian
$H_{{\rm eff}}$ of the modulated lattice in accordance with definition
$U(T)=\exp(-iH_{{\rm eff}}T).$ In contrast to instantaneous Hamiltonian
$H(\mathbf{k},t)$, the effective Hamiltonian $H_{{\rm eff}}$ is
$z$-independent, and it offers ``stroboscopic'' description of
the propagation dynamics over complete longitudinal period. The spectrum
of the system can be described by quasi-energies $\epsilon$ --- eigenvalues
of the effective Hamiltonian \cite{shirley.pr.138.B979.1965,sambe.pra.7.2203.1973}
--- that can be obtained from the expression $U(T)|\phi\rangle=\exp(-i\epsilon T)|\phi\rangle$.
Using this approach in the case of infinite discrete lattice we search
for solutions of Eq. (\ref{eq1}) in the form of periodic Bloch waves $A_{n}=A\exp(ikx_{n})$
and $B_{n}=B\exp(ikx_{n}+ika)$, where $x_{n}=2na$ is the discrete
transverse coordinate, and $k\in[-\pi/2a,\pi/2a]$ is the Bloch momentum
in the first Brillouin zone.
Substituting these expressions into Eq. (\ref{eq1}), one obtains
\begin{equation}
\begin{split}i\frac{dA}{dz}= & [J_{1}(z)\exp(iak_{x})+J_{2}(z)\exp(-iak_{x})]B,\\
i\frac{dB}{dz}= & [J_{1}(z)\exp(-iak_{x})+J_{2}(z)\exp(iak_{x})]A.
\end{split}
\label{eq2}
\end{equation}

\begin{figure}[htpb]
\centering \includegraphics[width=0.5\columnwidth]{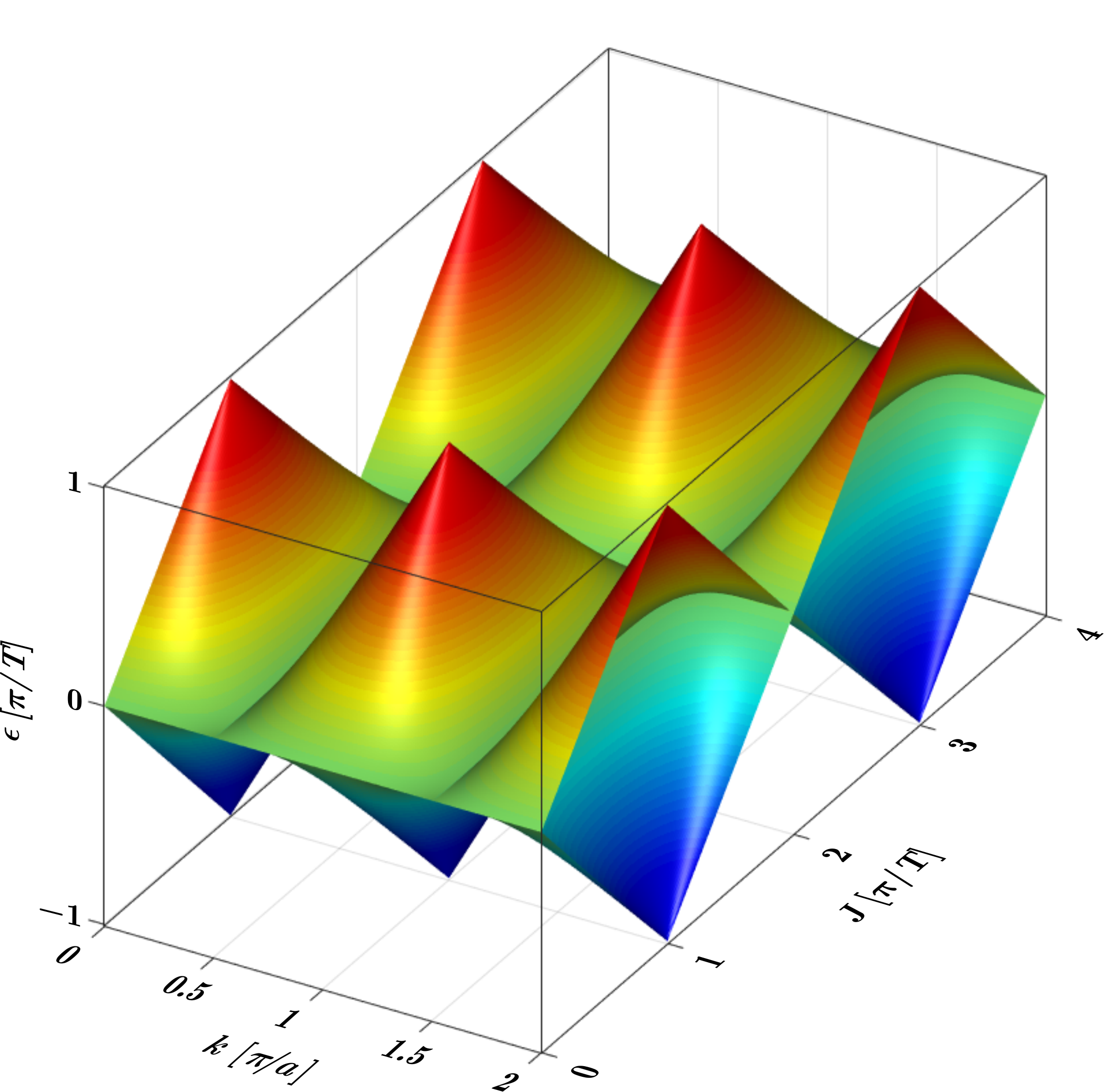}
\caption{Quasi-energy as a function of Bloch momentum $k$ and coupling constant
$J$.}
\label{fig2}
\end{figure}

Thus, Floquet evolution operator over one period can be represented
as\cite{kitagawa.prb.82.235114.2010,rudner.prx.3.031005.2013,maczewsky.nc.8.13756.2017}
\begin{align}
 & U(T)=\exp(-iH_{2}T/2)\exp(-iH_{1}T/2)\label{eq3}\\
 & =\cos(ak)\exp(-iak)\times\notag\\
 & \begin{bmatrix}\cos(JT)+i\tan(ak) & i\exp(iak)\sin(JT)\\
i\exp(iak)\sin(JT) & \exp(2iak)[\cos(JT)-i\tan(ak)]
\end{bmatrix},\notag
\end{align}
where Hamiltonians on the first and second half-periods are given
by
\[
\begin{split}H_{1}=\begin{bmatrix}0 & J\exp(iak)\\
J\exp(-iak) & 0
\end{bmatrix},\\
H_{2}=\begin{bmatrix}0 & J\exp(-iak)\\
J\exp(iak) & 0
\end{bmatrix}.
\end{split}
\]

\begin{figure}[htpb]
\centering \includegraphics[width=0.5\columnwidth]{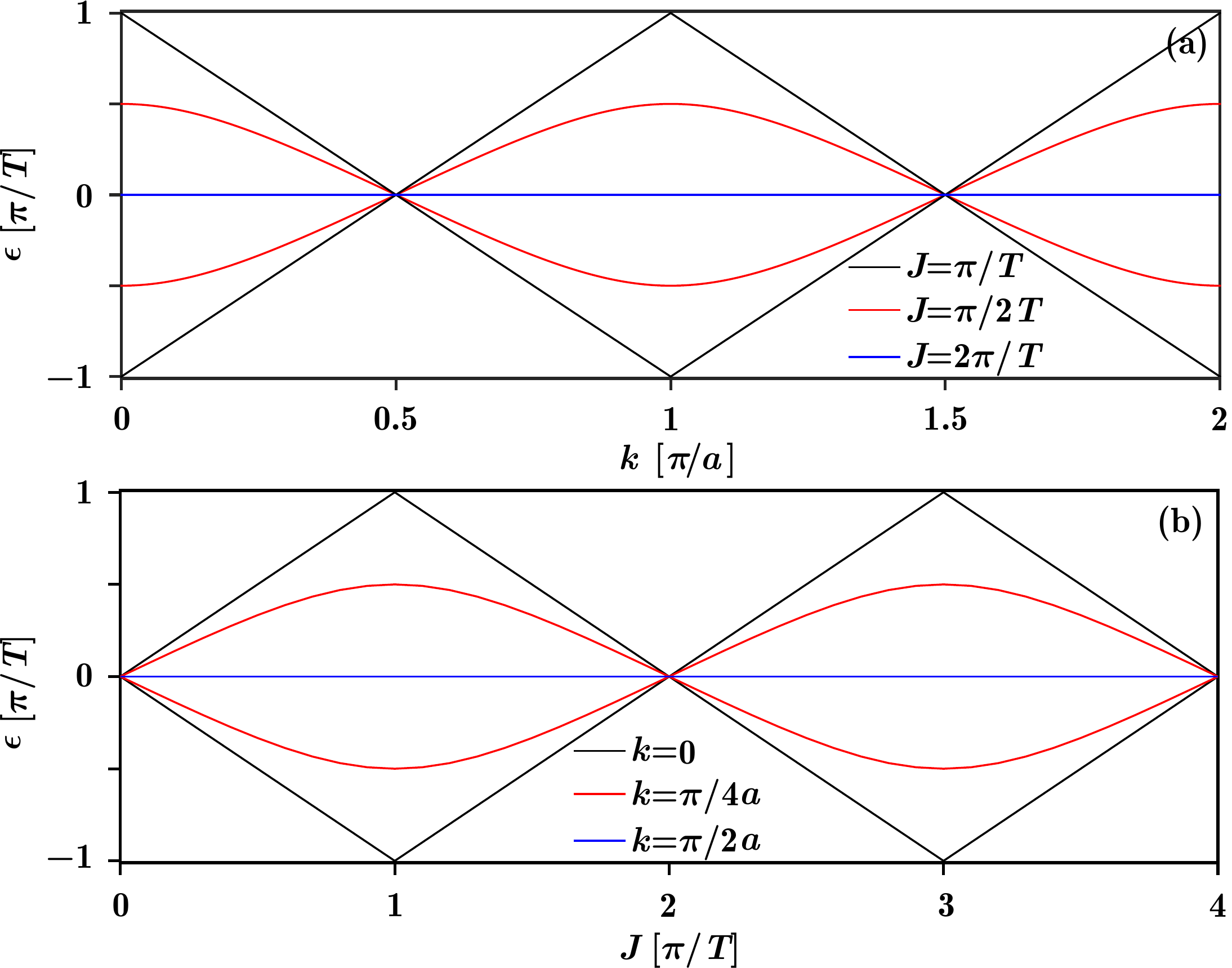}
\caption{(a) Quasi-energy as a function of $k$ for different $J$ values.
(b) Quasi-energy as a function of $J$ for different $k$ values.}
\label{fig3}
\end{figure}

One can see from Eq. (\ref{eq3}) that Floquet evolution operator
is a periodic function of transverse momentum $k$ with a period $\pi/a$
and of the coupling strength $J$ with a period $2\pi/T$ . Similarly,
by introducing the effective Hamiltonian through $U=\exp(-iH_{{\rm eff}}T)$
and calculating its eigenvalues (quasi-energies $\epsilon$), one
obtains that the latter are also periodic functions of $k$ and $J$.
In Fig. \ref{fig2}, we depict the dependence $\epsilon(k,J)$. Quasi-energy
band is symmetric with respect to the plane $\epsilon=0$ (it
is periodic also in the vertical direction with a period $2\pi/T$
because eigenvalues of periodic system are defined modulo $2\pi/T$
). The maxima of quasi-energies within vertical interval shown in
Fig. 2 are located at $k=n\pi/a$ and $J=(2l+1)\pi/T$, where $n$
is an integer and $l$ is a non-negative integer. To highlight the
details of this dependence we show quasi-energies in Figs. \ref{fig3}(a)
and \ref{fig3}(b) for certain fixed values of coupling strength $J$
and Bloch momentum $k$, respectively. Importantly, it follows from
Fig. \ref{fig3}(a) that the quasi-energy band is dispersive at $J<\pi/T$
(see red curves), so for this coupling strength any localized wavepacket
launched into system will diffract. When $J$ increases up to $\pi/T$
the dependence $\epsilon(k)$ becomes linear \cite{valle.pra.87.022119.2013}
(see black lines). This means that effective dispersion coefficient
vanishes and excitations in such a lattice will propagate without
diffraction, but with nonzero transverse velocity --- this is the rectification
regime. Further increase of the coupling strength makes quasi-energy
band dispersive again. Finally, quasi-energy band collapses to a line
at $J=2\pi/T$ (see the blue line). In this regime of dynamical localization
the shape of any wavepacket launched into system will be exactly reproduced
after one longitudinal period. Very similar transformations can be
observed for different Bloch momenta, when quasi-energy is plotted
as a function of coupling constant $J$, as shown in Fig. \ref{fig3}(b).

\begin{figure}[htpb]
\centering \includegraphics[width=0.5\columnwidth]{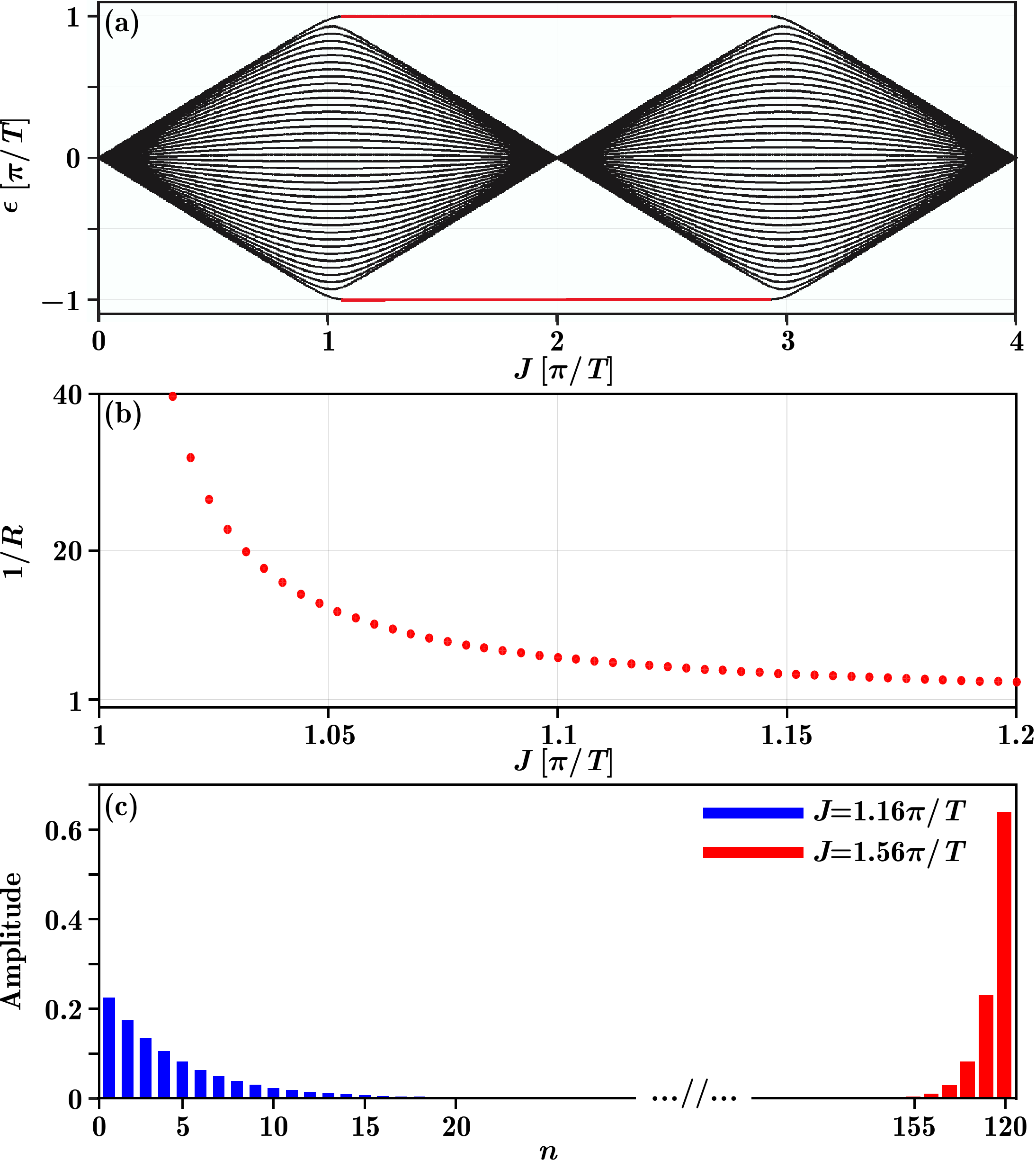} \caption{(a) Dependence of quasi-energies on the coupling constant in the finite
superlattice containing 200 sites in each sublattice. (b) Width of
the edge state versus coupling constant. (c) Absolute value of the edge states corresponding
to $J=1.16\pi/T$ and $J=1.56\pi/T$, respectively. }
\label{fig4}
\end{figure}

The situation changes qualitatively when the superlattice is truncated
in the transverse direction. In this case one cannot introduce Bloch
momentum anymore, so evolution dynamics is described by the system
(\ref{eq1}), where equations for amplitudes in the edge sites $A_{1}$ and
$A_{N}$ are replaced by the equations
$
i{dA_{1}}/{dz}=J_{1}(z)B_{1},~i{dA_{N}}/{dz}=J_{2}(z)B_{N-1}.
$
One should stress that the properties of the system do not change
qualitatively if superlattice is truncated on the site belonging to
sublattice $A$ on the left side, and on the site belonging to sublattice
$B$ on the right side. By introducing effective Hamiltonian for the
finite longitudinally modulated superlattice, one can determine its
quasi-energies that can be plotted as a function of the coupling strength
$J$. In Fig. \ref{fig4}(a) we display corresponding dependence.
One can see that this dependence inherits some features of $\epsilon(J)$
dependence of the infinite lattice [compare Figs. \ref{fig4}(a) and \ref{fig3}(b)].
Among them is the (partial) collapse of the quasi-energy band for
specific values of the coupling constant $J=2\pi m/T$. At the same
time, there are two qualitative differences. First, within the interval
$J\in[\pi/T,~3\pi/T]$ of coupling constants the isolated quasi-energies
emerged (see red lines) that are associated with edge states. In fact,
such edge states appear periodically in the intervals $[(4m+1)\pi/T,~(4m+3)\pi/T]$,
where $m$ is an integer. The second difference is that the period
of the $\epsilon(J)$ dependence is doubled in comparison with dependence
in the infinite lattice. Qualitative modification of the quasi-energy
spectrum indicates on the topological transition that
occurs in finite modulated superlattice upon variation of coupling
strength between waveguides. Interestingly, the collapse of quasi-energy
band at $J=2\pi/T$ indicating on the presence of dynamic localization
in the system coexists with the fact of formation of edge states,
so for this particular value of $J$ two qualitative different localization
mechanisms are simultaneously available.

The width of emerging edge states strongly depends on the coupling
constant. To illustrate this we introduce the participation ratio
$R=\sum_{n}|q_{n}|^{4}/(\sum_{n}|q_{n}|^{2})^{2}$, where $q_{n}=A_{n},B_{n}$
stands for light amplitudes on sites of sublattices $A$ and $B$. The width
of the mode is inversely proportional to participation ratio. In Fig.
\ref{fig4}(b), we show the width of the edge state versus coupling
constant $J$. Localization increases with increase of coupling constant,
so that already at $J>1.1\pi/T$ the edge state occupies less than
ten sites of the lattice. Maximal localization in nearly single surface
channel occurs at $J=2\pi/T$, and further increase of the coupling
constant leads to gradual delocalization of the edge state. Examples
of profiles of edge states (absolute value) with notably different localization degrees
at $J=1.16\pi/T$ and $J=1.56\pi/T$ are shown in
Fig. \ref{fig4}(c) .

\section{Transport properties}

\begin{figure*}[htpb]
\centering \includegraphics[width=1\textwidth]{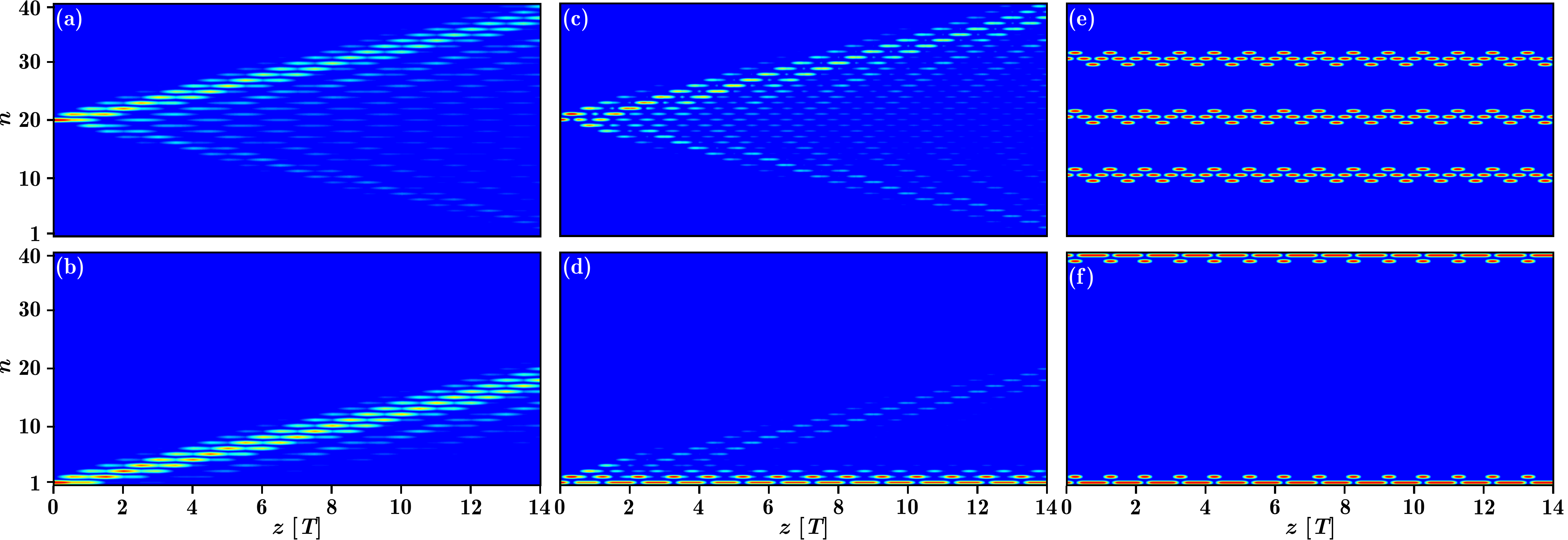} \caption{Propagation dynamics in the finite superlattice when only one site
in sublattice $A$ is excited by the beam $\exp[-{16\ln2}\,(x-x_{A})^{2}]$,
where $x_A$ is the coordinate of the site in sublattice $A$. Left
panels: internal site is excited. Right panels: edge site is excited.
First column: $J=0.5\pi/T$. Second column: $J=1.5\pi/T$.
Third column: $J=2\pi/T$. Parameters: $a=1$ and $T=1$.}
\label{fig5}
\end{figure*}

Topological transition that occurs in finite longitudinally modulated
superlattice suggests the existence of novel propagation scenarios
in this system. To study transport properties in such structures we
simultaneously consider excitations of the internal and edge sites
and use three representative values of the coupling constant. In the
particular realization of the lattice, that we use to study propagation
dynamics (see Fig. \ref{fig5}) two edge sites belong to different sublattices,
i.e. the ``bottom'' site belongs to sublattice $A$, while the ``top''
site belongs to sublattice $B$. First, we consider the case $J=0.5\pi/T$,
where quasi-energy band has finite width, while edge states do not
appear. In Fig. \ref{fig5}(a), we excite the internal waveguide and
find that the beam diffracts during propagation. Similarly, the excitation
of the edge waveguide shown in Fig. \ref{fig5}(b) is also accompanied
by rapid diffraction without any signatures of localization. Second,
we turn to the system with the coupling strength $J=1.5\pi/T$. For
this coupling constant according to Fig. \ref{fig4}(a), the width
of the quasi-energy band is still finite, but edge states already
emerge. Therefore, if an internal site is excited, discrete diffraction
will be observed as shown in Fig. \ref{fig5}(c). In contrast, excitation
of the edge site leads to the formation of well-localized edge state
and only weak radiation can be detected, as shown in Fig. \ref{fig5}(d).
The reason for small radiation is that we use excitation that does
not match directly the shape of the edge state, hence delocalized
bulk modes are excited too, but with small weights. Finally, we consider
the case with $J=2\pi/T$, where quasi-energy band collapses {[}Fig.
\ref{fig4}(a){]}. In this particular case dynamic localization occurs
irrespectively of the location of the excited site. In Fig. \ref{fig5}(e),
we show such a localization for excitations of sites number 10, 20,
and 30. In addition, we also excite the edge waveguides in Fig. \ref{fig5}(f),
where one can see that light beam does not experience expansion and
remains confined in two near-surface sites. This is the regime where
two distinct localization mechanisms coexist.

\begin{figure*}[htpb]
\centering \includegraphics[width=1\textwidth]{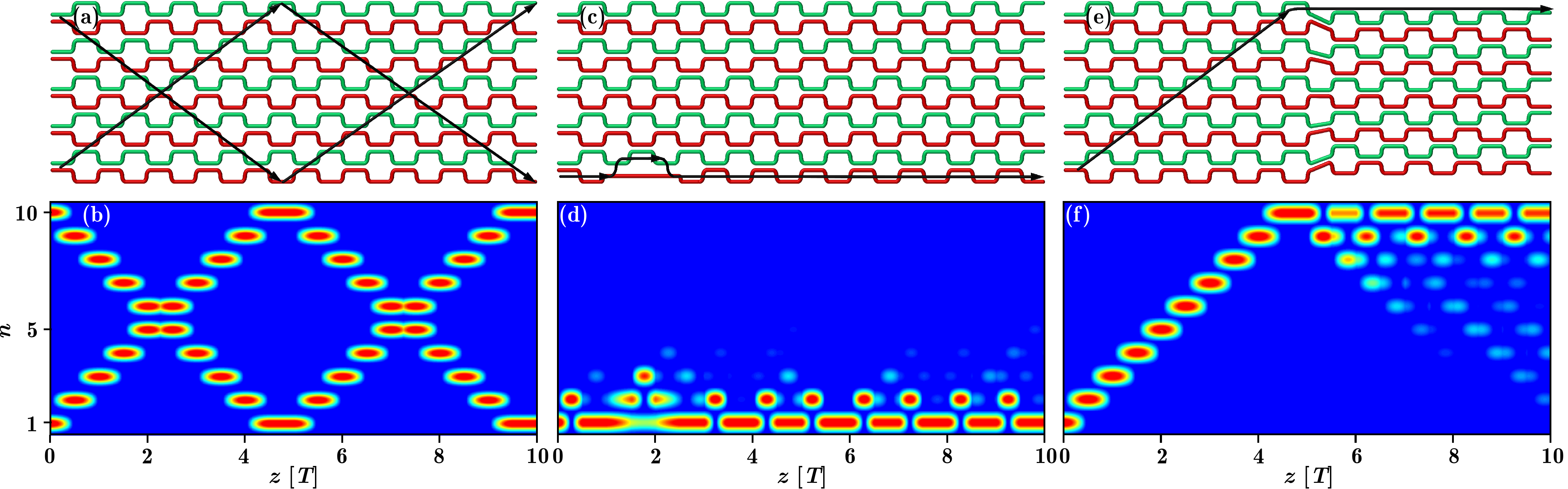}
\caption{(a) Schematic illustration of the waveguide array without defects
or deformation. Black curve with arrows indicate the propagation direction
for excitations of two opposite edge waveguides. (b) Propagation dynamics
in the lattice without defects at $J=\pi/T$. (c) and (d) Same as
(a) and (b), but for the system with a defect on the edge. For the
internal waveguides, the coupling strength is $J=1.8\pi/T$, while
for the edge straight waveguide the coupling strength is $J=\pi/4T$
at $T<z\le1.5T$ and $2T<z\le2.5T$, and $J=\pi/15T$ at $1.5T<z\le2T$.
(e) and (f) Same as (c) and (d), but for the lattice with global deformation
that changes coupling constants after certain distance $z$. The coupling
strength is $J=\pi/T$ at $0\le z<5T$ and $J=1.5\pi/T$ at $5T\le z<10T$.}
\label{fig6}
\end{figure*}

The propagation dynamics in this system is specific at $J=\pi/T$
and it deserves separate discussion. In the infinite lattice this
coupling constant corresponds to linear dependence of quasi-energy
on Bloch momentum $k$, i.e. the absence of diffraction (rectification
regime). Finite superlattice inherits this property to some extent,
i.e. localized excitations in finite lattice also do not diffract,
but move with constant transverse velocity. Interestingly, despite
the absence of diffraction, the excitation of edge states in this
regime does not occur, since moving excitations just change their
propagation direction when they hit edge sites. This is illustrated
in Figs. \ref{fig6}(a) and \ref{fig6}(b), where we simultaneously
excite two opposite edge waveguides. In this particular case we excited
sites belonging to different sublattices, as before, but dynamics
does not change qualitatively if sites from one sublattice are excited.
Notice that in this interesting regime the transverse confinement
occurs without any nonlinearity, and at the same time the propagation
trajectory of the beam and its output position can be flexibly controlled
that is advantageous for practical applications. To illustrate enhanced
robustness of edge states introduced here we deliberately introduce
considerable deformation at the surface of the lattice, by replacing
the whole section of the edge waveguide between $z=T$ and $z=2.5T$
with a straight section, as shown schematically in Fig. \ref{fig6}(c).
The coupling constant for internal waveguides is selected as $J=1.8\pi/T$,
i.e. it corresponds to situation when edge states form at the surface.
The corresponding propagation dynamics in this deformed structure
is shown in Fig. \ref{fig6}(d). Despite considerable deformation
of the structure the edge excitation passes the defect without noticable
scattering into the bulk of the lattice. However, it should be mentioned
that if surface defect is too long and extends over three or more
periods of the structure, the edge state may be destroyed and light
will penetrate into the depth of the lattice. Finally, we design a
structure that is composed of two parts with different coupling strengths
between waveguides: in the first part of the lattice $J=\pi/T$ for
closely spaced waveguides, while in the second part of the lattice
$J=1.5\pi/T$. Such variation in the coupling strength can be achieved
by reduction of the transverse period at certain distance $z$, as
shown in Fig. \ref{fig6}(e). Since in the first part of the lattice
the coupling constant is selected such that no edge states can form,
but diffractionless propagation is possible, the input beam will propagate
from one edge of the lattice towards opposite edge. If it arrives
to opposite edge in the point, where coupling constant changes and
edge states become possible, the beam may excite the edge state and
stay near the surface of the structure, as shown in Fig. \ref{fig6}(f).
If, however, the beam hits the opposite edge before the point where
coupling constant increases, it will be bounced back and enter into
right half of the lattice in one of the internal waveguides. This
will lead to fast diffraction of the beam without excitation of edge
states. This setting can be considered as a kind of optical switch,
where the presence of signal in the output edge channel depends on
the position of the input excitation.

\section{Conclusions}

Summarizing, we investigated transport properties in the one-dimensional
dynamical superlattices. We have shown that in finite modulated superlattices
topological transition may occur that leads to appearance of edge
states, whose degree of localization depends on the coupling constant
between lattice sites. This localization mechanism may coexist with
dynamic localization due to collapse of quasi-energy bands.

\section*{Acknowledgement}
This work was supported by China Postdoctoral Science Foundation (2016M600777,
2016M600776, 2016M590935), the National Natural Science Foundation
of China (11474228, 61605154), and Qatar National Research Fund (NPRP
6-021-1-005, 8-028-1-001).


\bibliographystyle{myprx}
\bibliography{my_refs_library}

\begin{thebibliography}{47}%
\makeatletter
\providecommand \@ifxundefined [1]{%
 \@ifx{#1\undefined}
}%
\providecommand \@ifnum [1]{%
 \ifnum #1\expandafter \@firstoftwo
 \else \expandafter \@secondoftwo
 \fi
}%
\providecommand \@ifx [1]{%
 \ifx #1\expandafter \@firstoftwo
 \else \expandafter \@secondoftwo
 \fi
}%
\providecommand \natexlab [1]{#1}%
\providecommand \enquote  [1]{``#1''}%
\providecommand \bibnamefont  [1]{#1}%
\providecommand \bibfnamefont [1]{#1}%
\providecommand \citenamefont [1]{#1}%
\providecommand \href@noop [0]{\@secondoftwo}%
\providecommand \href [0]{\begingroup \@sanitize@url \@href}%
\providecommand \@href[1]{\@@startlink{#1}\@@href}%
\providecommand \@@href[1]{\endgroup#1\@@endlink}%
\providecommand \@sanitize@url [0]{\catcode `\\12\catcode `\$12\catcode
  `\&12\catcode `\#12\catcode `\^12\catcode `\_12\catcode `\%12\relax}%
\providecommand \@@startlink[1]{}%
\providecommand \@@endlink[0]{}%
\providecommand \url  [0]{\begingroup\@sanitize@url \@url }%
\providecommand \@url [1]{\endgroup\@href {#1}{\urlprefix }}%
\providecommand \urlprefix  [0]{URL }%
\providecommand \Eprint [0]{\href }%
\providecommand \doibase [0]{http://dx.doi.org/}%
\providecommand \selectlanguage [0]{\@gobble}%
\providecommand \bibinfo  [0]{\@secondoftwo}%
\providecommand \bibfield  [0]{\@secondoftwo}%
\providecommand \translation [1]{[#1]}%
\providecommand \BibitemOpen [0]{}%
\providecommand \bibitemStop [0]{}%
\providecommand \bibitemNoStop [0]{.\EOS\space}%
\providecommand \EOS [0]{\spacefactor3000\relax}%
\providecommand \BibitemShut  [1]{\csname bibitem#1\endcsname}%
\let\auto@bib@innerbib\@empty
\bibitem [{\citenamefont {Oka}\ and\ \citenamefont
  {Aoki}(2009)}]{oka.prb.79.081406.2009}%
  \BibitemOpen
  \bibfield  {author} {\bibinfo {author} {\bibfnamefont {T.}~\bibnamefont
  {Oka}}\ and\ \bibinfo {author} {\bibfnamefont {H.}~\bibnamefont {Aoki}},\
  }\emph {\bibinfo {title} {Photovoltaic {H}all effect in graphene}},\ \href
  {\doibase 10.1103/PhysRevB.79.081406} {\bibfield  {journal} {\bibinfo
  {journal} {Phys. Rev. B}\ }\textbf {\bibinfo {volume} {79}},\ \bibinfo
  {pages} {081406} (\bibinfo {year} {2009})}\BibitemShut {NoStop}%
\bibitem [{\citenamefont {Kitagawa}\ \emph {et~al.}(2010)\citenamefont
  {Kitagawa}, \citenamefont {Berg}, \citenamefont {Rudner},\ and\ \citenamefont
  {Demler}}]{kitagawa.prb.82.235114.2010}%
  \BibitemOpen
  \bibfield  {author} {\bibinfo {author} {\bibfnamefont {T.}~\bibnamefont
  {Kitagawa}}, \bibinfo {author} {\bibfnamefont {E.}~\bibnamefont {Berg}},
  \bibinfo {author} {\bibfnamefont {M.}~\bibnamefont {Rudner}}, \ and\ \bibinfo
  {author} {\bibfnamefont {E.}~\bibnamefont {Demler}},\ }\emph {\bibinfo
  {title} {Topological characterization of periodically driven quantum
  systems}},\ \href {\doibase 10.1103/PhysRevB.82.235114} {\bibfield  {journal}
  {\bibinfo  {journal} {Phys. Rev. B}\ }\textbf {\bibinfo {volume} {82}},\
  \bibinfo {pages} {235114} (\bibinfo {year} {2010})}\BibitemShut {NoStop}%
\bibitem [{\citenamefont {Lindner}\ \emph {et~al.}(2011)\citenamefont
  {Lindner}, \citenamefont {Refael},\ and\ \citenamefont
  {Galitski}}]{lindner.np.7.490.2011}%
  \BibitemOpen
  \bibfield  {author} {\bibinfo {author} {\bibfnamefont {N.~H.}\ \bibnamefont
  {Lindner}}, \bibinfo {author} {\bibfnamefont {G.}~\bibnamefont {Refael}}, \
  and\ \bibinfo {author} {\bibfnamefont {V.}~\bibnamefont {Galitski}},\ }\emph
  {\bibinfo {title} {Floquet topological insulator in semiconductor quantum
  wells}},\ \href {\doibase 10.1038/nphys1926} {\bibfield  {journal} {\bibinfo
  {journal} {Nat. Phys.}\ }\textbf {\bibinfo {volume} {7}},\ \bibinfo {pages}
  {490} (\bibinfo {year} {2011})}\BibitemShut {NoStop}%
\bibitem [{\citenamefont {Rudner}\ \emph {et~al.}(2013)\citenamefont {Rudner},
  \citenamefont {Lindner}, \citenamefont {Berg},\ and\ \citenamefont
  {Levin}}]{rudner.prx.3.031005.2013}%
  \BibitemOpen
  \bibfield  {author} {\bibinfo {author} {\bibfnamefont {M.~S.}\ \bibnamefont
  {Rudner}}, \bibinfo {author} {\bibfnamefont {N.~H.}\ \bibnamefont {Lindner}},
  \bibinfo {author} {\bibfnamefont {E.}~\bibnamefont {Berg}}, \ and\ \bibinfo
  {author} {\bibfnamefont {M.}~\bibnamefont {Levin}},\ }\emph {\bibinfo {title}
  {Anomalous Edge States and the Bulk-Edge Correspondence for Periodically
  Driven Two-Dimensional Systems}},\ \href {\doibase 10.1103/PhysRevX.3.031005}
  {\bibfield  {journal} {\bibinfo  {journal} {Phys. Rev. X}\ }\textbf {\bibinfo
  {volume} {3}},\ \bibinfo {pages} {031005} (\bibinfo {year}
  {2013})}\BibitemShut {NoStop}%
\bibitem [{\citenamefont {G\'omez-Le\'on}\ and\ \citenamefont
  {Platero}(2013)}]{leon.prl.110.200403.2013}%
  \BibitemOpen
  \bibfield  {author} {\bibinfo {author} {\bibfnamefont {A.}~\bibnamefont
  {G\'omez-Le\'on}}\ and\ \bibinfo {author} {\bibfnamefont {G.}~\bibnamefont
  {Platero}},\ }\emph {\bibinfo {title} {Floquet-{B}loch Theory and Topology in
  Periodically Driven Lattices}},\ \href {\doibase
  10.1103/PhysRevLett.110.200403} {\bibfield  {journal} {\bibinfo  {journal}
  {Phys. Rev. Lett.}\ }\textbf {\bibinfo {volume} {110}},\ \bibinfo {pages}
  {200403} (\bibinfo {year} {2013})}\BibitemShut {NoStop}%
\bibitem [{\citenamefont {Goldman}\ and\ \citenamefont
  {Dalibard}(2014)}]{goldman.prx.4.031027.2014}%
  \BibitemOpen
  \bibfield  {author} {\bibinfo {author} {\bibfnamefont {N.}~\bibnamefont
  {Goldman}}\ and\ \bibinfo {author} {\bibfnamefont {J.}~\bibnamefont
  {Dalibard}},\ }\emph {\bibinfo {title} {Periodically Driven Quantum Systems:
  Effective Hamiltonians and Engineered Gauge Fields}},\ \href {\doibase
  10.1103/PhysRevX.4.031027} {\bibfield  {journal} {\bibinfo  {journal} {Phys.
  Rev. X}\ }\textbf {\bibinfo {volume} {4}},\ \bibinfo {pages} {031027}
  (\bibinfo {year} {2014})}\BibitemShut {NoStop}%
\bibitem [{\citenamefont {Asboth}\ and\ \citenamefont
  {Edge}(2015)}]{asboth.pra.91.022324.2015}%
  \BibitemOpen
  \bibfield  {author} {\bibinfo {author} {\bibfnamefont {J.~K.}\ \bibnamefont
  {Asboth}}\ and\ \bibinfo {author} {\bibfnamefont {J.~M.}\ \bibnamefont
  {Edge}},\ }\emph {\bibinfo {title} {Edge-state-enhanced transport in a
  two-dimensional quantum walk}},\ \href {\doibase 10.1103/PhysRevA.91.022324}
  {\bibfield  {journal} {\bibinfo  {journal} {Phys. Rev. A}\ }\textbf {\bibinfo
  {volume} {91}},\ \bibinfo {pages} {022324} (\bibinfo {year}
  {2015})}\BibitemShut {NoStop}%
\bibitem [{\citenamefont {Xiong}\ \emph {et~al.}(2016)\citenamefont {Xiong},
  \citenamefont {Gong},\ and\ \citenamefont {An}}]{xiong.prb.93.184306.2016}%
  \BibitemOpen
  \bibfield  {author} {\bibinfo {author} {\bibfnamefont {T.-S.}\ \bibnamefont
  {Xiong}}, \bibinfo {author} {\bibfnamefont {J.}~\bibnamefont {Gong}}, \ and\
  \bibinfo {author} {\bibfnamefont {J.-H.}\ \bibnamefont {An}},\ }\emph
  {\bibinfo {title} {Towards large-{C}hern-number topological phases by
  periodic quenching}},\ \href {\doibase 10.1103/PhysRevB.93.184306} {\bibfield
   {journal} {\bibinfo  {journal} {Phys. Rev. B}\ }\textbf {\bibinfo {volume}
  {93}},\ \bibinfo {pages} {184306} (\bibinfo {year} {2016})}\BibitemShut
  {NoStop}%
\bibitem [{\citenamefont {Titum}\ \emph {et~al.}(2016)\citenamefont {Titum},
  \citenamefont {Berg}, \citenamefont {Rudner}, \citenamefont {Refael},\ and\
  \citenamefont {Lindner}}]{titum.prx.6.021013.2016}%
  \BibitemOpen
  \bibfield  {author} {\bibinfo {author} {\bibfnamefont {P.}~\bibnamefont
  {Titum}}, \bibinfo {author} {\bibfnamefont {E.}~\bibnamefont {Berg}},
  \bibinfo {author} {\bibfnamefont {M.~S.}\ \bibnamefont {Rudner}}, \bibinfo
  {author} {\bibfnamefont {G.}~\bibnamefont {Refael}}, \ and\ \bibinfo {author}
  {\bibfnamefont {N.~H.}\ \bibnamefont {Lindner}},\ }\emph {\bibinfo {title}
  {Anomalous {Floquet-Anderson} Insulator as a Nonadiabatic Quantized Charge
  Pump}},\ \href {\doibase 10.1103/PhysRevX.6.021013} {\bibfield  {journal}
  {\bibinfo  {journal} {Phys. Rev. X}\ }\textbf {\bibinfo {volume} {6}},\
  \bibinfo {pages} {021013} (\bibinfo {year} {2016})}\BibitemShut {NoStop}%
\bibitem [{\citenamefont {Yu}\ and\ \citenamefont
  {Fan}(2009)}]{yu.np.3.91.2009}%
  \BibitemOpen
  \bibfield  {author} {\bibinfo {author} {\bibfnamefont {Z.}~\bibnamefont
  {Yu}}\ and\ \bibinfo {author} {\bibfnamefont {S.}~\bibnamefont {Fan}},\
  }\emph {\bibinfo {title} {Complete optical isolation created by indirect
  interband photonic transitions}},\ \href {\doibase 10.1038/nphoton.2008.273}
  {\bibfield  {journal} {\bibinfo  {journal} {Nat. Photon.}\ }\textbf {\bibinfo
  {volume} {3}},\ \bibinfo {pages} {91} (\bibinfo {year} {2009})}\BibitemShut
  {NoStop}%
\bibitem [{\citenamefont {Fang}\ \emph {et~al.}(2012)\citenamefont {Fang},
  \citenamefont {Yu},\ and\ \citenamefont {Fan}}]{fang.np.6.782.2012}%
  \BibitemOpen
  \bibfield  {author} {\bibinfo {author} {\bibfnamefont {K.}~\bibnamefont
  {Fang}}, \bibinfo {author} {\bibfnamefont {Z.}~\bibnamefont {Yu}}, \ and\
  \bibinfo {author} {\bibfnamefont {S.}~\bibnamefont {Fan}},\ }\emph {\bibinfo
  {title} {Realizing effective magnetic field for photons by controlling the
  phase of dynamic modulation}},\ \href {\doibase 10.1038/nphoton.2012.236}
  {\bibfield  {journal} {\bibinfo  {journal} {Nat. Photon.}\ }\textbf {\bibinfo
  {volume} {6}},\ \bibinfo {pages} {782} (\bibinfo {year} {2012})}\BibitemShut
  {NoStop}%
\bibitem [{\citenamefont {Kitagawa}\ \emph {et~al.}(2012)\citenamefont
  {Kitagawa}, \citenamefont {Broome}, \citenamefont {Fedrizzi}, \citenamefont
  {Rudner}, \citenamefont {Berg}, \citenamefont {Kassal}, \citenamefont
  {Aspuru-Guzik}, \citenamefont {Demler},\ and\ \citenamefont
  {White}}]{kitagawa.nc.3.882.2012}%
  \BibitemOpen
  \bibfield  {author} {\bibinfo {author} {\bibfnamefont {T.}~\bibnamefont
  {Kitagawa}}, \bibinfo {author} {\bibfnamefont {M.~A.}\ \bibnamefont
  {Broome}}, \bibinfo {author} {\bibfnamefont {A.}~\bibnamefont {Fedrizzi}},
  \bibinfo {author} {\bibfnamefont {M.~S.}\ \bibnamefont {Rudner}}, \bibinfo
  {author} {\bibfnamefont {E.}~\bibnamefont {Berg}}, \bibinfo {author}
  {\bibfnamefont {I.}~\bibnamefont {Kassal}}, \bibinfo {author} {\bibfnamefont
  {A.}~\bibnamefont {Aspuru-Guzik}}, \bibinfo {author} {\bibfnamefont
  {E.}~\bibnamefont {Demler}}, \ and\ \bibinfo {author} {\bibfnamefont {A.~G.}\
  \bibnamefont {White}},\ }\emph {\bibinfo {title} {Observation of
  topologically protected bound states in photonic quantum walks}},\ \href
  {\doibase 10.1038/ncomms1872} {\bibfield  {journal} {\bibinfo  {journal}
  {Nat. Commun.}\ }\textbf {\bibinfo {volume} {3}},\ \bibinfo {pages} {882}
  (\bibinfo {year} {2012})}\BibitemShut {NoStop}%
\bibitem [{\citenamefont {Khanikaev}\ \emph {et~al.}(2012)\citenamefont
  {Khanikaev}, \citenamefont {Mousavi}, \citenamefont {Tse}, \citenamefont
  {Kargarian}, \citenamefont {MacDonald},\ and\ \citenamefont
  {Shvets}}]{khanikaev.nm.12.233.2012}%
  \BibitemOpen
  \bibfield  {author} {\bibinfo {author} {\bibfnamefont {A.~B.}\ \bibnamefont
  {Khanikaev}}, \bibinfo {author} {\bibfnamefont {S.~H.}\ \bibnamefont
  {Mousavi}}, \bibinfo {author} {\bibfnamefont {W.-K.}\ \bibnamefont {Tse}},
  \bibinfo {author} {\bibfnamefont {M.}~\bibnamefont {Kargarian}}, \bibinfo
  {author} {\bibfnamefont {A.~H.}\ \bibnamefont {MacDonald}}, \ and\ \bibinfo
  {author} {\bibfnamefont {G.}~\bibnamefont {Shvets}},\ }\emph {\bibinfo
  {title} {Photonic topological insulators}},\ \href {\doibase
  10.1038/nmat3520} {\bibfield  {journal} {\bibinfo  {journal} {Nat. Mater.}\
  }\textbf {\bibinfo {volume} {12}},\ \bibinfo {pages} {233} (\bibinfo {year}
  {2012})}\BibitemShut {NoStop}%
\bibitem [{\citenamefont {Liang}\ and\ \citenamefont
  {Chong}(2013)}]{liang.prl.110.203904.2013}%
  \BibitemOpen
  \bibfield  {author} {\bibinfo {author} {\bibfnamefont {G.~Q.}\ \bibnamefont
  {Liang}}\ and\ \bibinfo {author} {\bibfnamefont {Y.~D.}\ \bibnamefont
  {Chong}},\ }\emph {\bibinfo {title} {Optical Resonator Analog of a
  Two-Dimensional Topological Insulator}},\ \href {\doibase
  10.1103/PhysRevLett.110.203904} {\bibfield  {journal} {\bibinfo  {journal}
  {Phys. Rev. Lett.}\ }\textbf {\bibinfo {volume} {110}},\ \bibinfo {pages}
  {203904} (\bibinfo {year} {2013})}\BibitemShut {NoStop}%
\bibitem [{\citenamefont {Rechtsman}\ \emph {et~al.}(2013)\citenamefont
  {Rechtsman}, \citenamefont {Zeuner}, \citenamefont {Plotnik}, \citenamefont
  {Lumer}, \citenamefont {Podolsky}, \citenamefont {Dreisow}, \citenamefont
  {Nolte}, \citenamefont {Segev},\ and\ \citenamefont
  {Szameit}}]{rechtsman.nature.496.196.2013}%
  \BibitemOpen
  \bibfield  {author} {\bibinfo {author} {\bibfnamefont {M.~C.}\ \bibnamefont
  {Rechtsman}}, \bibinfo {author} {\bibfnamefont {J.~M.}\ \bibnamefont
  {Zeuner}}, \bibinfo {author} {\bibfnamefont {Y.}~\bibnamefont {Plotnik}},
  \bibinfo {author} {\bibfnamefont {Y.}~\bibnamefont {Lumer}}, \bibinfo
  {author} {\bibfnamefont {D.}~\bibnamefont {Podolsky}}, \bibinfo {author}
  {\bibfnamefont {F.}~\bibnamefont {Dreisow}}, \bibinfo {author} {\bibfnamefont
  {S.}~\bibnamefont {Nolte}}, \bibinfo {author} {\bibfnamefont
  {M.}~\bibnamefont {Segev}}, \ and\ \bibinfo {author} {\bibfnamefont
  {A.}~\bibnamefont {Szameit}},\ }\emph {\bibinfo {title} {Photonic {F}loquet
  topological insulators}},\ \href {\doibase 10.1038/nature12066} {\bibfield
  {journal} {\bibinfo  {journal} {Nature}\ }\textbf {\bibinfo {volume} {496}},\
  \bibinfo {pages} {196} (\bibinfo {year} {2013})}\BibitemShut {NoStop}%
\bibitem [{\citenamefont {Chen}\ \emph {et~al.}(2014)\citenamefont {Chen},
  \citenamefont {Jiang}, \citenamefont {Chen}, \citenamefont {Zhu},
  \citenamefont {Zhou}, \citenamefont {Dong},\ and\ \citenamefont
  {Chan}}]{chen.nc.5.5782.2014}%
  \BibitemOpen
  \bibfield  {author} {\bibinfo {author} {\bibfnamefont {W.-J.}\ \bibnamefont
  {Chen}}, \bibinfo {author} {\bibfnamefont {S.-J.}\ \bibnamefont {Jiang}},
  \bibinfo {author} {\bibfnamefont {X.-D.}\ \bibnamefont {Chen}}, \bibinfo
  {author} {\bibfnamefont {B.}~\bibnamefont {Zhu}}, \bibinfo {author}
  {\bibfnamefont {L.}~\bibnamefont {Zhou}}, \bibinfo {author} {\bibfnamefont
  {J.-W.}\ \bibnamefont {Dong}}, \ and\ \bibinfo {author} {\bibfnamefont
  {C.~T.}\ \bibnamefont {Chan}},\ }\emph {\bibinfo {title} {Experimental
  realization of photonic topological insulator in a uniaxial metacrystal
  waveguide}},\ \href {\doibase 10.1038/ncomms6782} {\bibfield  {journal}
  {\bibinfo  {journal} {Nat. Commun.}\ }\textbf {\bibinfo {volume} {5}},\
  \bibinfo {pages} {5782} (\bibinfo {year} {2014})}\BibitemShut {NoStop}%
\bibitem [{\citenamefont {Leykam}\ \emph {et~al.}(2016)\citenamefont {Leykam},
  \citenamefont {Rechtsman},\ and\ \citenamefont
  {Chong}}]{leykam.prl.117.013902.2016}%
  \BibitemOpen
  \bibfield  {author} {\bibinfo {author} {\bibfnamefont {D.}~\bibnamefont
  {Leykam}}, \bibinfo {author} {\bibfnamefont {M.~C.}\ \bibnamefont
  {Rechtsman}}, \ and\ \bibinfo {author} {\bibfnamefont {Y.~D.}\ \bibnamefont
  {Chong}},\ }\emph {\bibinfo {title} {Anomalous Topological Phases and
  Unpaired {D}irac Cones in Photonic {F}loquet Topological Insulators}},\ \href
  {\doibase 10.1103/PhysRevLett.117.013902} {\bibfield  {journal} {\bibinfo
  {journal} {Phys. Rev. Lett.}\ }\textbf {\bibinfo {volume} {117}},\ \bibinfo
  {pages} {013902} (\bibinfo {year} {2016})}\BibitemShut {NoStop}%
\bibitem [{\citenamefont {Leykam}\ and\ \citenamefont
  {Chong}(2016)}]{leykam.prl.117.143901.2016}%
  \BibitemOpen
  \bibfield  {author} {\bibinfo {author} {\bibfnamefont {D.}~\bibnamefont
  {Leykam}}\ and\ \bibinfo {author} {\bibfnamefont {Y.~D.}\ \bibnamefont
  {Chong}},\ }\emph {\bibinfo {title} {Edge Solitons in Nonlinear-Photonic
  Topological Insulators}},\ \href {\doibase 10.1103/PhysRevLett.117.143901}
  {\bibfield  {journal} {\bibinfo  {journal} {Phys. Rev. Lett.}\ }\textbf
  {\bibinfo {volume} {117}},\ \bibinfo {pages} {143901} (\bibinfo {year}
  {2016})}\BibitemShut {NoStop}%
\bibitem [{\citenamefont {Weimann}\ \emph {et~al.}(2017)\citenamefont
  {Weimann}, \citenamefont {Kremer}, \citenamefont {Plotnik}, \citenamefont
  {Lumer}, \citenamefont {Nolte}, \citenamefont {Makris}, \citenamefont
  {Segev}, \citenamefont {Rechtsman},\ and\ \citenamefont
  {Szameit}}]{weimann.nm.16.1476.2017}%
  \BibitemOpen
  \bibfield  {author} {\bibinfo {author} {\bibfnamefont {S.}~\bibnamefont
  {Weimann}}, \bibinfo {author} {\bibfnamefont {M.}~\bibnamefont {Kremer}},
  \bibinfo {author} {\bibfnamefont {Y.}~\bibnamefont {Plotnik}}, \bibinfo
  {author} {\bibfnamefont {Y.}~\bibnamefont {Lumer}}, \bibinfo {author}
  {\bibfnamefont {S.}~\bibnamefont {Nolte}}, \bibinfo {author} {\bibfnamefont
  {K.~G.}\ \bibnamefont {Makris}}, \bibinfo {author} {\bibfnamefont
  {M.}~\bibnamefont {Segev}}, \bibinfo {author} {\bibfnamefont {M.~C.}\
  \bibnamefont {Rechtsman}}, \ and\ \bibinfo {author} {\bibfnamefont
  {A.}~\bibnamefont {Szameit}},\ }\emph {\bibinfo {title} {Topologically
  protected bound states in photonic parity-time-symmetric crystals}},\ \href
  {\doibase 10.1038/nmat4811} {\bibfield  {journal} {\bibinfo  {journal} {Nat.
  Mater.}\ }\textbf {\bibinfo {volume} {16}},\ \bibinfo {pages} {433} (\bibinfo
  {year} {2017})}\BibitemShut {NoStop}%
\bibitem [{\citenamefont {Longhi}(2009{\natexlab{a}})}]{longhi.lpr.3.243.2009}%
  \BibitemOpen
  \bibfield  {author} {\bibinfo {author} {\bibfnamefont {S.}~\bibnamefont
  {Longhi}},\ }\emph {\bibinfo {title} {Quantum-optical analogies using
  photonic structures}},\ \href {\doibase 10.1002/lpor.200810055} {\bibfield
  {journal} {\bibinfo  {journal} {Laser Photon. Rev.}\ }\textbf {\bibinfo
  {volume} {3}},\ \bibinfo {pages} {243} (\bibinfo {year}
  {2009}{\natexlab{a}})}\BibitemShut {NoStop}%
\bibitem [{\citenamefont {Garanovich}\ \emph {et~al.}(2012)\citenamefont
  {Garanovich}, \citenamefont {Longhi}, \citenamefont {Sukhorukov},\ and\
  \citenamefont {Kivshar}}]{garanovich.pr.518.1.2012}%
  \BibitemOpen
  \bibfield  {author} {\bibinfo {author} {\bibfnamefont {I.~L.}\ \bibnamefont
  {Garanovich}}, \bibinfo {author} {\bibfnamefont {S.}~\bibnamefont {Longhi}},
  \bibinfo {author} {\bibfnamefont {A.~A.}\ \bibnamefont {Sukhorukov}}, \ and\
  \bibinfo {author} {\bibfnamefont {Y.~S.}\ \bibnamefont {Kivshar}},\ }\emph
  {\bibinfo {title} {Light propagation and localization in modulated photonic
  lattices and waveguides}},\ \href {\doibase 10.1016/j.physrep.2012.03.005}
  {\bibfield  {journal} {\bibinfo  {journal} {Phys. Rep.}\ }\textbf {\bibinfo
  {volume} {518}},\ \bibinfo {pages} {1} (\bibinfo {year} {2012})}\BibitemShut
  {NoStop}%
\bibitem [{\citenamefont {Zhang}\ \emph {et~al.}(2015)\citenamefont {Zhang},
  \citenamefont {Wu}, \citenamefont {Beli\'c}, \citenamefont {Zheng},
  \citenamefont {Wang}, \citenamefont {Xiao},\ and\ \citenamefont
  {Zhang}}]{zhang.lpr.9.331.2015}%
  \BibitemOpen
  \bibfield  {author} {\bibinfo {author} {\bibfnamefont {Y.~Q.}\ \bibnamefont
  {Zhang}}, \bibinfo {author} {\bibfnamefont {Z.~K.}\ \bibnamefont {Wu}},
  \bibinfo {author} {\bibfnamefont {M.~R.}\ \bibnamefont {Beli\'c}}, \bibinfo
  {author} {\bibfnamefont {H.~B.}\ \bibnamefont {Zheng}}, \bibinfo {author}
  {\bibfnamefont {Z.~G.}\ \bibnamefont {Wang}}, \bibinfo {author}
  {\bibfnamefont {M.}~\bibnamefont {Xiao}}, \ and\ \bibinfo {author}
  {\bibfnamefont {Y.~P.}\ \bibnamefont {Zhang}},\ }\emph {\bibinfo {title}
  {Photonic {F}loquet topological insulators in atomic ensembles}},\ \href
  {\doibase 10.1002/lpor.201400428} {\bibfield  {journal} {\bibinfo  {journal}
  {Laser Photon. Rev.}\ }\textbf {\bibinfo {volume} {9}},\ \bibinfo {pages}
  {331} (\bibinfo {year} {2015})}\BibitemShut {NoStop}%
\bibitem [{\citenamefont {Maczewsky}\ \emph {et~al.}(2017)\citenamefont
  {Maczewsky}, \citenamefont {Zeuner}, \citenamefont {Nolte},\ and\
  \citenamefont {Szameit}}]{maczewsky.nc.8.13756.2017}%
  \BibitemOpen
  \bibfield  {author} {\bibinfo {author} {\bibfnamefont {L.~J.}\ \bibnamefont
  {Maczewsky}}, \bibinfo {author} {\bibfnamefont {J.~M.}\ \bibnamefont
  {Zeuner}}, \bibinfo {author} {\bibfnamefont {S.}~\bibnamefont {Nolte}}, \
  and\ \bibinfo {author} {\bibfnamefont {A.}~\bibnamefont {Szameit}},\ }\emph
  {\bibinfo {title} {Observation of photonic anomalous {F}loquet topological
  insulators}},\ \href {\doibase 10.1038/ncomms13756} {\bibfield  {journal}
  {\bibinfo  {journal} {Nat. Commun.}\ }\textbf {\bibinfo {volume} {8}},\
  \bibinfo {pages} {13756} (\bibinfo {year} {2017})}\BibitemShut {NoStop}%
\bibitem [{\citenamefont {Broome}\ \emph {et~al.}(2010)\citenamefont {Broome},
  \citenamefont {Fedrizzi}, \citenamefont {Lanyon}, \citenamefont {Kassal},
  \citenamefont {Aspuru-Guzik},\ and\ \citenamefont
  {White}}]{broome.prl.104.153602.2010}%
  \BibitemOpen
  \bibfield  {author} {\bibinfo {author} {\bibfnamefont {M.~A.}\ \bibnamefont
  {Broome}}, \bibinfo {author} {\bibfnamefont {A.}~\bibnamefont {Fedrizzi}},
  \bibinfo {author} {\bibfnamefont {B.~P.}\ \bibnamefont {Lanyon}}, \bibinfo
  {author} {\bibfnamefont {I.}~\bibnamefont {Kassal}}, \bibinfo {author}
  {\bibfnamefont {A.}~\bibnamefont {Aspuru-Guzik}}, \ and\ \bibinfo {author}
  {\bibfnamefont {A.~G.}\ \bibnamefont {White}},\ }\emph {\bibinfo {title}
  {Discrete Single-Photon Quantum Walks with Tunable Decoherence}},\ \href
  {\doibase 10.1103/PhysRevLett.104.153602} {\bibfield  {journal} {\bibinfo
  {journal} {Phys. Rev. Lett.}\ }\textbf {\bibinfo {volume} {104}},\ \bibinfo
  {pages} {153602} (\bibinfo {year} {2010})}\BibitemShut {NoStop}%
\bibitem [{\citenamefont {Sansoni}\ \emph {et~al.}(2012)\citenamefont
  {Sansoni}, \citenamefont {Sciarrino}, \citenamefont {Vallone}, \citenamefont
  {Mataloni}, \citenamefont {Crespi}, \citenamefont {Ramponi},\ and\
  \citenamefont {Osellame}}]{sansoni.prl.108.010502.2012}%
  \BibitemOpen
  \bibfield  {author} {\bibinfo {author} {\bibfnamefont {L.}~\bibnamefont
  {Sansoni}}, \bibinfo {author} {\bibfnamefont {F.}~\bibnamefont {Sciarrino}},
  \bibinfo {author} {\bibfnamefont {G.}~\bibnamefont {Vallone}}, \bibinfo
  {author} {\bibfnamefont {P.}~\bibnamefont {Mataloni}}, \bibinfo {author}
  {\bibfnamefont {A.}~\bibnamefont {Crespi}}, \bibinfo {author} {\bibfnamefont
  {R.}~\bibnamefont {Ramponi}}, \ and\ \bibinfo {author} {\bibfnamefont
  {R.}~\bibnamefont {Osellame}},\ }\emph {\bibinfo {title} {Two-Particle
  {Bosonic-Fermionic} Quantum Walk via Integrated Photonics}},\ \href {\doibase
  10.1103/PhysRevLett.108.010502} {\bibfield  {journal} {\bibinfo  {journal}
  {Phys. Rev. Lett.}\ }\textbf {\bibinfo {volume} {108}},\ \bibinfo {pages}
  {010502} (\bibinfo {year} {2012})}\BibitemShut {NoStop}%
\bibitem [{\citenamefont {Asb\'oth}\ \emph {et~al.}(2014)\citenamefont
  {Asb\'oth}, \citenamefont {Tarasinski},\ and\ \citenamefont
  {Delplace}}]{asboth.prb.90.125143.2014}%
  \BibitemOpen
  \bibfield  {author} {\bibinfo {author} {\bibfnamefont {J.~K.}\ \bibnamefont
  {Asb\'oth}}, \bibinfo {author} {\bibfnamefont {B.}~\bibnamefont
  {Tarasinski}}, \ and\ \bibinfo {author} {\bibfnamefont {P.}~\bibnamefont
  {Delplace}},\ }\emph {\bibinfo {title} {Chiral symmetry and bulk-boundary
  correspondence in periodically driven one-dimensional systems}},\ \href
  {\doibase 10.1103/PhysRevB.90.125143} {\bibfield  {journal} {\bibinfo
  {journal} {Phys. Rev. B}\ }\textbf {\bibinfo {volume} {90}},\ \bibinfo
  {pages} {125143} (\bibinfo {year} {2014})}\BibitemShut {NoStop}%
\bibitem [{\citenamefont {Dal~Lago}\ \emph {et~al.}(2015)\citenamefont
  {Dal~Lago}, \citenamefont {Atala},\ and\ \citenamefont
  {Foa~Torres}}]{lago.pra.92.023624.2015}%
  \BibitemOpen
  \bibfield  {author} {\bibinfo {author} {\bibfnamefont {V.}~\bibnamefont
  {Dal~Lago}}, \bibinfo {author} {\bibfnamefont {M.}~\bibnamefont {Atala}}, \
  and\ \bibinfo {author} {\bibfnamefont {L.~E.~F.}\ \bibnamefont
  {Foa~Torres}},\ }\emph {\bibinfo {title} {Floquet topological transitions in
  a driven one-dimensional topological insulator}},\ \href {\doibase
  10.1103/PhysRevA.92.023624} {\bibfield  {journal} {\bibinfo  {journal} {Phys.
  Rev. A}\ }\textbf {\bibinfo {volume} {92}},\ \bibinfo {pages} {023624}
  (\bibinfo {year} {2015})}\BibitemShut {NoStop}%
\bibitem [{\citenamefont {Longhi}\ \emph {et~al.}(2006)\citenamefont {Longhi},
  \citenamefont {Marangoni}, \citenamefont {Lobino}, \citenamefont {Ramponi},
  \citenamefont {Laporta}, \citenamefont {Cianci},\ and\ \citenamefont
  {Foglietti}}]{longhi.prl.96.243901.2006}%
  \BibitemOpen
  \bibfield  {author} {\bibinfo {author} {\bibfnamefont {S.}~\bibnamefont
  {Longhi}}, \bibinfo {author} {\bibfnamefont {M.}~\bibnamefont {Marangoni}},
  \bibinfo {author} {\bibfnamefont {M.}~\bibnamefont {Lobino}}, \bibinfo
  {author} {\bibfnamefont {R.}~\bibnamefont {Ramponi}}, \bibinfo {author}
  {\bibfnamefont {P.}~\bibnamefont {Laporta}}, \bibinfo {author} {\bibfnamefont
  {E.}~\bibnamefont {Cianci}}, \ and\ \bibinfo {author} {\bibfnamefont
  {V.}~\bibnamefont {Foglietti}},\ }\emph {\bibinfo {title} {Observation of
  Dynamic Localization in Periodically Curved Waveguide Arrays}},\ \href
  {\doibase 10.1103/PhysRevLett.96.243901} {\bibfield  {journal} {\bibinfo
  {journal} {Phys. Rev. Lett.}\ }\textbf {\bibinfo {volume} {96}},\ \bibinfo
  {pages} {243901} (\bibinfo {year} {2006})}\BibitemShut {NoStop}%
\bibitem [{\citenamefont {Garanovich}\ \emph {et~al.}(2008)\citenamefont
  {Garanovich}, \citenamefont {Sukhorukov},\ and\ \citenamefont
  {Kivshar}}]{garanovich.prl.100.203904.2008}%
  \BibitemOpen
  \bibfield  {author} {\bibinfo {author} {\bibfnamefont {I.~L.}\ \bibnamefont
  {Garanovich}}, \bibinfo {author} {\bibfnamefont {A.~A.}\ \bibnamefont
  {Sukhorukov}}, \ and\ \bibinfo {author} {\bibfnamefont {Y.~S.}\ \bibnamefont
  {Kivshar}},\ }\emph {\bibinfo {title} {Defect-Free Surface States in
  Modulated Photonic Lattices}},\ \href {\doibase
  10.1103/PhysRevLett.100.203904} {\bibfield  {journal} {\bibinfo  {journal}
  {Phys. Rev. Lett.}\ }\textbf {\bibinfo {volume} {100}},\ \bibinfo {pages}
  {203904} (\bibinfo {year} {2008})}\BibitemShut {NoStop}%
\bibitem [{\citenamefont {Szameit}\ \emph {et~al.}(2008)\citenamefont
  {Szameit}, \citenamefont {Garanovich}, \citenamefont {Heinrich},
  \citenamefont {Sukhorukov}, \citenamefont {Dreisow}, \citenamefont {Pertsch},
  \citenamefont {Nolte}, \citenamefont {T\"unnermann},\ and\ \citenamefont
  {Kivshar}}]{szameit.prl.101.203902.2008}%
  \BibitemOpen
  \bibfield  {author} {\bibinfo {author} {\bibfnamefont {A.}~\bibnamefont
  {Szameit}}, \bibinfo {author} {\bibfnamefont {I.~L.}\ \bibnamefont
  {Garanovich}}, \bibinfo {author} {\bibfnamefont {M.}~\bibnamefont
  {Heinrich}}, \bibinfo {author} {\bibfnamefont {A.~A.}\ \bibnamefont
  {Sukhorukov}}, \bibinfo {author} {\bibfnamefont {F.}~\bibnamefont {Dreisow}},
  \bibinfo {author} {\bibfnamefont {T.}~\bibnamefont {Pertsch}}, \bibinfo
  {author} {\bibfnamefont {S.}~\bibnamefont {Nolte}}, \bibinfo {author}
  {\bibfnamefont {A.}~\bibnamefont {T\"unnermann}}, \ and\ \bibinfo {author}
  {\bibfnamefont {Y.~S.}\ \bibnamefont {Kivshar}},\ }\emph {\bibinfo {title}
  {Observation of Defect-Free Surface Modes in Optical Waveguide Arrays}},\
  \href {\doibase 10.1103/PhysRevLett.101.203902} {\bibfield  {journal}
  {\bibinfo  {journal} {Phys. Rev. Lett.}\ }\textbf {\bibinfo {volume} {101}},\
  \bibinfo {pages} {203902} (\bibinfo {year} {2008})}\BibitemShut {NoStop}%
\bibitem [{\citenamefont {Longhi}\ and\ \citenamefont
  {Staliunas}(2008)}]{longhi.oc.281.4343.2008}%
  \BibitemOpen
  \bibfield  {author} {\bibinfo {author} {\bibfnamefont {S.}~\bibnamefont
  {Longhi}}\ and\ \bibinfo {author} {\bibfnamefont {K.}~\bibnamefont
  {Staliunas}},\ }\emph {\bibinfo {title} {Self-collimation and self-imaging
  effects in modulated waveguide arrays}},\ \href {\doibase
  10.1016/j.optcom.2008.05.014} {\bibfield  {journal} {\bibinfo  {journal}
  {Opt. Commun.}\ }\textbf {\bibinfo {volume} {281}},\ \bibinfo {pages} {4343}
  (\bibinfo {year} {2008})}\BibitemShut {NoStop}%
\bibitem [{\citenamefont {Szameit}\ \emph {et~al.}(2009)\citenamefont
  {Szameit}, \citenamefont {Kartashov}, \citenamefont {Dreisow}, \citenamefont
  {Heinrich}, \citenamefont {Pertsch}, \citenamefont {Nolte}, \citenamefont
  {T\"unnermann}, \citenamefont {Vysloukh}, \citenamefont {Lederer},\ and\
  \citenamefont {Torner}}]{szameit.prl.102.153901.2009}%
  \BibitemOpen
  \bibfield  {author} {\bibinfo {author} {\bibfnamefont {A.}~\bibnamefont
  {Szameit}}, \bibinfo {author} {\bibfnamefont {Y.~V.}\ \bibnamefont
  {Kartashov}}, \bibinfo {author} {\bibfnamefont {F.}~\bibnamefont {Dreisow}},
  \bibinfo {author} {\bibfnamefont {M.}~\bibnamefont {Heinrich}}, \bibinfo
  {author} {\bibfnamefont {T.}~\bibnamefont {Pertsch}}, \bibinfo {author}
  {\bibfnamefont {S.}~\bibnamefont {Nolte}}, \bibinfo {author} {\bibfnamefont
  {A.}~\bibnamefont {T\"unnermann}}, \bibinfo {author} {\bibfnamefont {V.~A.}\
  \bibnamefont {Vysloukh}}, \bibinfo {author} {\bibfnamefont {F.}~\bibnamefont
  {Lederer}}, \ and\ \bibinfo {author} {\bibfnamefont {L.}~\bibnamefont
  {Torner}},\ }\emph {\bibinfo {title} {Inhibition of Light Tunneling in
  Waveguide Arrays}},\ \href {\doibase 10.1103/PhysRevLett.102.153901}
  {\bibfield  {journal} {\bibinfo  {journal} {Phys. Rev. Lett.}\ }\textbf
  {\bibinfo {volume} {102}},\ \bibinfo {pages} {153901} (\bibinfo {year}
  {2009})}\BibitemShut {NoStop}%
\bibitem [{\citenamefont
  {Longhi}(2009{\natexlab{b}})}]{longhi.prb.80.235102.2009}%
  \BibitemOpen
  \bibfield  {author} {\bibinfo {author} {\bibfnamefont {S.}~\bibnamefont
  {Longhi}},\ }\emph {\bibinfo {title} {Dynamic localization and transport in
  complex crystals}},\ \href {\doibase 10.1103/PhysRevB.80.235102} {\bibfield
  {journal} {\bibinfo  {journal} {Phys. Rev. B}\ }\textbf {\bibinfo {volume}
  {80}},\ \bibinfo {pages} {235102} (\bibinfo {year}
  {2009}{\natexlab{b}})}\BibitemShut {NoStop}%
\bibitem [{\citenamefont {Kartashov}\ \emph {et~al.}(2009)\citenamefont
  {Kartashov}, \citenamefont {Szameit}, \citenamefont {Vysloukh},\ and\
  \citenamefont {Torner}}]{kartashov.ol.34.2906.2009}%
  \BibitemOpen
  \bibfield  {author} {\bibinfo {author} {\bibfnamefont {Y.~V.}\ \bibnamefont
  {Kartashov}}, \bibinfo {author} {\bibfnamefont {A.}~\bibnamefont {Szameit}},
  \bibinfo {author} {\bibfnamefont {V.~A.}\ \bibnamefont {Vysloukh}}, \ and\
  \bibinfo {author} {\bibfnamefont {L.}~\bibnamefont {Torner}},\ }\emph
  {\bibinfo {title} {Light tunneling inhibition and anisotropic diffraction
  engineering in two-dimensional waveguide arrays}},\ \href {\doibase
  10.1364/OL.34.002906} {\bibfield  {journal} {\bibinfo  {journal} {Opt.
  Lett.}\ }\textbf {\bibinfo {volume} {34}},\ \bibinfo {pages} {2906} (\bibinfo
  {year} {2009})}\BibitemShut {NoStop}%
\bibitem [{\citenamefont {Longhi}(2009{\natexlab{c}})}]{longhi.ol.34.458.2009}%
  \BibitemOpen
  \bibfield  {author} {\bibinfo {author} {\bibfnamefont {S.}~\bibnamefont
  {Longhi}},\ }\emph {\bibinfo {title} {Rectification of light refraction in
  curved waveguide arrays}},\ \href {\doibase 10.1364/OL.34.000458} {\bibfield
  {journal} {\bibinfo  {journal} {Opt. Lett.}\ }\textbf {\bibinfo {volume}
  {34}},\ \bibinfo {pages} {458} (\bibinfo {year}
  {2009}{\natexlab{c}})}\BibitemShut {NoStop}%
\bibitem [{\citenamefont {Dreisow}\ \emph {et~al.}(2013)\citenamefont
  {Dreisow}, \citenamefont {Kartashov}, \citenamefont {Heinrich}, \citenamefont
  {Vysloukh}, \citenamefont {T¨¹nnermann}, \citenamefont {Nolte}, \citenamefont
  {Torner}, \citenamefont {Longhi},\ and\ \citenamefont
  {Szameit}}]{dreisow.epl.101.44002.2013}%
  \BibitemOpen
  \bibfield  {author} {\bibinfo {author} {\bibfnamefont {F.}~\bibnamefont
  {Dreisow}}, \bibinfo {author} {\bibfnamefont {Y.~V.}\ \bibnamefont
  {Kartashov}}, \bibinfo {author} {\bibfnamefont {M.}~\bibnamefont {Heinrich}},
  \bibinfo {author} {\bibfnamefont {V.~A.}\ \bibnamefont {Vysloukh}}, \bibinfo
  {author} {\bibfnamefont {A.}~\bibnamefont {T¨¹nnermann}}, \bibinfo {author}
  {\bibfnamefont {S.}~\bibnamefont {Nolte}}, \bibinfo {author} {\bibfnamefont
  {L.}~\bibnamefont {Torner}}, \bibinfo {author} {\bibfnamefont
  {S.}~\bibnamefont {Longhi}}, \ and\ \bibinfo {author} {\bibfnamefont
  {A.}~\bibnamefont {Szameit}},\ }\emph {\bibinfo {title} {Spatial light
  rectification in an optical waveguide lattice}},\ \href {\doibase
  10.1209/0295-5075/101/44002} {\bibfield  {journal} {\bibinfo  {journal} {EPL
  (Europhys. Lett.)}\ }\textbf {\bibinfo {volume} {101}},\ \bibinfo {pages}
  {44002} (\bibinfo {year} {2013})}\BibitemShut {NoStop}%
\bibitem [{\citenamefont {Kartashov}\ \emph {et~al.}(2016)\citenamefont
  {Kartashov}, \citenamefont {Vysloukh}, \citenamefont {Konotop},\ and\
  \citenamefont {Torner}}]{kartashov.pra.93.013841.2016}%
  \BibitemOpen
  \bibfield  {author} {\bibinfo {author} {\bibfnamefont {Y.~V.}\ \bibnamefont
  {Kartashov}}, \bibinfo {author} {\bibfnamefont {V.~A.}\ \bibnamefont
  {Vysloukh}}, \bibinfo {author} {\bibfnamefont {V.~V.}\ \bibnamefont
  {Konotop}}, \ and\ \bibinfo {author} {\bibfnamefont {L.}~\bibnamefont
  {Torner}},\ }\emph {\bibinfo {title} {Diffraction control in
  $\mathcal{PT}$-symmetric photonic lattices: From beam rectification to
  dynamic localization}},\ \href {\doibase 10.1103/PhysRevA.93.013841}
  {\bibfield  {journal} {\bibinfo  {journal} {Phys. Rev. A}\ }\textbf {\bibinfo
  {volume} {93}},\ \bibinfo {pages} {013841} (\bibinfo {year}
  {2016})}\BibitemShut {NoStop}%
\bibitem [{\citenamefont {Asb{\'o}th}\ \emph {et~al.}(2016)\citenamefont
  {Asb{\'o}th}, \citenamefont {Oroszl{\'a}ny},\ and\ \citenamefont
  {P{\'a}lyi}}]{asboth.book.2016}%
  \BibitemOpen
  \bibfield  {author} {\bibinfo {author} {\bibfnamefont {J.~K.}\ \bibnamefont
  {Asb{\'o}th}}, \bibinfo {author} {\bibfnamefont {L.}~\bibnamefont
  {Oroszl{\'a}ny}}, \ and\ \bibinfo {author} {\bibfnamefont {A.}~\bibnamefont
  {P{\'a}lyi}},\ }\emph {\bibinfo {title} {The {Su-Schrieffer-Heeger} (SSH)
  Model}},\ in\ \href {\doibase 10.1007/978-3-319-25607-8_1} {\emph {\bibinfo
  {booktitle} {A Short Course on Topological Insulators: Band Structure and
  Edge States in One and Two Dimensions}}}\ (\bibinfo  {publisher} {Springer},\
  \bibinfo {address} {Cham},\ \bibinfo {year} {2016})\ pp.\ \bibinfo {pages}
  {1--22}\BibitemShut {NoStop}%
\bibitem [{\citenamefont {Plotnik}\ \emph {et~al.}(2014)\citenamefont
  {Plotnik}, \citenamefont {Rechtsman}, \citenamefont {Song}, \citenamefont
  {Heinrich}, \citenamefont {Zeuner}, \citenamefont {Nolte}, \citenamefont
  {Lumer}, \citenamefont {Malkova}, \citenamefont {Xu}, \citenamefont
  {Szameit}, \citenamefont {Chen},\ and\ \citenamefont
  {Segev}}]{plotnik.nm.13.57.2014}%
  \BibitemOpen
  \bibfield  {author} {\bibinfo {author} {\bibfnamefont {Y.}~\bibnamefont
  {Plotnik}}, \bibinfo {author} {\bibfnamefont {M.~C.}\ \bibnamefont
  {Rechtsman}}, \bibinfo {author} {\bibfnamefont {D.}~\bibnamefont {Song}},
  \bibinfo {author} {\bibfnamefont {M.}~\bibnamefont {Heinrich}}, \bibinfo
  {author} {\bibfnamefont {J.~M.}\ \bibnamefont {Zeuner}}, \bibinfo {author}
  {\bibfnamefont {S.}~\bibnamefont {Nolte}}, \bibinfo {author} {\bibfnamefont
  {Y.}~\bibnamefont {Lumer}}, \bibinfo {author} {\bibfnamefont
  {N.}~\bibnamefont {Malkova}}, \bibinfo {author} {\bibfnamefont
  {J.}~\bibnamefont {Xu}}, \bibinfo {author} {\bibfnamefont {A.}~\bibnamefont
  {Szameit}}, \bibinfo {author} {\bibfnamefont {Z.}~\bibnamefont {Chen}}, \
  and\ \bibinfo {author} {\bibfnamefont {M.}~\bibnamefont {Segev}},\ }\emph
  {\bibinfo {title} {Observation of unconventional edge states in `photonic
  graphene'}},\ \href {\doibase 10.1038/nmat3783} {\bibfield  {journal}
  {\bibinfo  {journal} {Nat. Mater.}\ }\textbf {\bibinfo {volume} {13}},\
  \bibinfo {pages} {57} (\bibinfo {year} {2014})}\BibitemShut {NoStop}%
\bibitem [{\citenamefont {Vicencio}\ \emph {et~al.}(2015)\citenamefont
  {Vicencio}, \citenamefont {Cantillano}, \citenamefont {Morales-Inostroza},
  \citenamefont {Real}, \citenamefont {Mej\'{i}a-Cort\'es}, \citenamefont
  {Weimann}, \citenamefont {Szameit},\ and\ \citenamefont
  {Molina}}]{vicencio.prl.114.245503.2015}%
  \BibitemOpen
  \bibfield  {author} {\bibinfo {author} {\bibfnamefont {R.~A.}\ \bibnamefont
  {Vicencio}}, \bibinfo {author} {\bibfnamefont {C.}~\bibnamefont
  {Cantillano}}, \bibinfo {author} {\bibfnamefont {L.}~\bibnamefont
  {Morales-Inostroza}}, \bibinfo {author} {\bibfnamefont {B.}~\bibnamefont
  {Real}}, \bibinfo {author} {\bibfnamefont {C.}~\bibnamefont
  {Mej\'{i}a-Cort\'es}}, \bibinfo {author} {\bibfnamefont {S.}~\bibnamefont
  {Weimann}}, \bibinfo {author} {\bibfnamefont {A.}~\bibnamefont {Szameit}}, \
  and\ \bibinfo {author} {\bibfnamefont {M.~I.}\ \bibnamefont {Molina}},\
  }\emph {\bibinfo {title} {Observation of Localized States in {L}ieb Photonic
  Lattices}},\ \href {\doibase 10.1103/PhysRevLett.114.245503} {\bibfield
  {journal} {\bibinfo  {journal} {Phys. Rev. Lett.}\ }\textbf {\bibinfo
  {volume} {114}},\ \bibinfo {pages} {245503} (\bibinfo {year}
  {2015})}\BibitemShut {NoStop}%
\bibitem [{\citenamefont {Mukherjee}\ \emph {et~al.}(2015)\citenamefont
  {Mukherjee}, \citenamefont {Spracklen}, \citenamefont {Choudhury},
  \citenamefont {Goldman}, \citenamefont {\"Ohberg}, \citenamefont
  {Andersson},\ and\ \citenamefont {Thomson}}]{mukherjee.prl.114.245504.2015}%
  \BibitemOpen
  \bibfield  {author} {\bibinfo {author} {\bibfnamefont {S.}~\bibnamefont
  {Mukherjee}}, \bibinfo {author} {\bibfnamefont {A.}~\bibnamefont
  {Spracklen}}, \bibinfo {author} {\bibfnamefont {D.}~\bibnamefont
  {Choudhury}}, \bibinfo {author} {\bibfnamefont {N.}~\bibnamefont {Goldman}},
  \bibinfo {author} {\bibfnamefont {P.}~\bibnamefont {\"Ohberg}}, \bibinfo
  {author} {\bibfnamefont {E.}~\bibnamefont {Andersson}}, \ and\ \bibinfo
  {author} {\bibfnamefont {R.~R.}\ \bibnamefont {Thomson}},\ }\emph {\bibinfo
  {title} {Observation of a Localized Flat-Band State in a Photonic {L}ieb
  Lattice}},\ \href {\doibase 10.1103/PhysRevLett.114.245504} {\bibfield
  {journal} {\bibinfo  {journal} {Phys. Rev. Lett.}\ }\textbf {\bibinfo
  {volume} {114}},\ \bibinfo {pages} {245504} (\bibinfo {year}
  {2015})}\BibitemShut {NoStop}%
\bibitem [{\citenamefont {Diebel}\ \emph {et~al.}(2016)\citenamefont {Diebel},
  \citenamefont {Leykam}, \citenamefont {Kroesen}, \citenamefont {Denz},\ and\
  \citenamefont {Desyatnikov}}]{diebel.prl.116.183902.2016}%
  \BibitemOpen
  \bibfield  {author} {\bibinfo {author} {\bibfnamefont {F.}~\bibnamefont
  {Diebel}}, \bibinfo {author} {\bibfnamefont {D.}~\bibnamefont {Leykam}},
  \bibinfo {author} {\bibfnamefont {S.}~\bibnamefont {Kroesen}}, \bibinfo
  {author} {\bibfnamefont {C.}~\bibnamefont {Denz}}, \ and\ \bibinfo {author}
  {\bibfnamefont {A.~S.}\ \bibnamefont {Desyatnikov}},\ }\emph {\bibinfo
  {title} {Conical Diffraction and Composite {L}ieb Bosons in Photonic
  Lattices}},\ \href {\doibase 10.1103/PhysRevLett.116.183902} {\bibfield
  {journal} {\bibinfo  {journal} {Phys. Rev. Lett.}\ }\textbf {\bibinfo
  {volume} {116}},\ \bibinfo {pages} {183902} (\bibinfo {year}
  {2016})}\BibitemShut {NoStop}%
\bibitem [{\citenamefont {Szameit}\ \emph {et~al.}(2011)\citenamefont
  {Szameit}, \citenamefont {Rechtsman}, \citenamefont {Bahat-Treidel},\ and\
  \citenamefont {Segev}}]{szameit.pra.84.021806.2011}%
  \BibitemOpen
  \bibfield  {author} {\bibinfo {author} {\bibfnamefont {A.}~\bibnamefont
  {Szameit}}, \bibinfo {author} {\bibfnamefont {M.~C.}\ \bibnamefont
  {Rechtsman}}, \bibinfo {author} {\bibfnamefont {O.}~\bibnamefont
  {Bahat-Treidel}}, \ and\ \bibinfo {author} {\bibfnamefont {M.}~\bibnamefont
  {Segev}},\ }\emph {\bibinfo {title} {$\mathcal{PT}$-symmetry in honeycomb
  photonic lattices}},\ \href {\doibase 10.1103/PhysRevA.84.021806} {\bibfield
  {journal} {\bibinfo  {journal} {Phys. Rev. A}\ }\textbf {\bibinfo {volume}
  {84}},\ \bibinfo {pages} {021806} (\bibinfo {year} {2011})}\BibitemShut
  {NoStop}%
\bibitem [{\citenamefont {Weimann}\ \emph {et~al.}(2016)\citenamefont
  {Weimann}, \citenamefont {Morales-Inostroza}, \citenamefont {Real},
  \citenamefont {Cantillano}, \citenamefont {Szameit},\ and\ \citenamefont
  {Vicencio}}]{weimann.ol.41.2414.2016}%
  \BibitemOpen
  \bibfield  {author} {\bibinfo {author} {\bibfnamefont {S.}~\bibnamefont
  {Weimann}}, \bibinfo {author} {\bibfnamefont {L.}~\bibnamefont
  {Morales-Inostroza}}, \bibinfo {author} {\bibfnamefont {B.}~\bibnamefont
  {Real}}, \bibinfo {author} {\bibfnamefont {C.}~\bibnamefont {Cantillano}},
  \bibinfo {author} {\bibfnamefont {A.}~\bibnamefont {Szameit}}, \ and\
  \bibinfo {author} {\bibfnamefont {R.~A.}\ \bibnamefont {Vicencio}},\ }\emph
  {\bibinfo {title} {Transport in {Sawtooth} photonic lattices}},\ \href
  {\doibase 10.1364/OL.41.002414} {\bibfield  {journal} {\bibinfo  {journal}
  {Opt. Lett.}\ }\textbf {\bibinfo {volume} {41}},\ \bibinfo {pages} {2414}
  (\bibinfo {year} {2016})}\BibitemShut {NoStop}%
\bibitem [{\citenamefont {Shirley}(1965)}]{shirley.pr.138.B979.1965}%
  \BibitemOpen
  \bibfield  {author} {\bibinfo {author} {\bibfnamefont {J.~H.}\ \bibnamefont
  {Shirley}},\ }\emph {\bibinfo {title} {Solution of the {S}chr\"odinger
  Equation with a Hamiltonian Periodic in Time}},\ \href {\doibase
  10.1103/PhysRev.138.B979} {\bibfield  {journal} {\bibinfo  {journal} {Phys.
  Rev.}\ }\textbf {\bibinfo {volume} {138}},\ \bibinfo {pages} {B979} (\bibinfo
  {year} {1965})}\BibitemShut {NoStop}%
\bibitem [{\citenamefont {Sambe}(1973)}]{sambe.pra.7.2203.1973}%
  \BibitemOpen
  \bibfield  {author} {\bibinfo {author} {\bibfnamefont {H.}~\bibnamefont
  {Sambe}},\ }\emph {\bibinfo {title} {Steady States and Quasienergies of a
  Quantum-Mechanical System in an Oscillating Field}},\ \href {\doibase
  10.1103/PhysRevA.7.2203} {\bibfield  {journal} {\bibinfo  {journal} {Phys.
  Rev. A}\ }\textbf {\bibinfo {volume} {7}},\ \bibinfo {pages} {2203} (\bibinfo
  {year} {1973})}\BibitemShut {NoStop}%
\bibitem [{\citenamefont {Della~Valle}\ and\ \citenamefont
  {Longhi}(2013)}]{valle.pra.87.022119.2013}%
  \BibitemOpen
  \bibfield  {author} {\bibinfo {author} {\bibfnamefont {G.}~\bibnamefont
  {Della~Valle}}\ and\ \bibinfo {author} {\bibfnamefont {S.}~\bibnamefont
  {Longhi}},\ }\emph {\bibinfo {title} {Spectral and transport properties of
  time-periodic $\mathcal{PT}$-symmetric tight-binding lattices}},\ \href
  {\doibase 10.1103/PhysRevA.87.022119} {\bibfield  {journal} {\bibinfo
  {journal} {Phys. Rev. A}\ }\textbf {\bibinfo {volume} {87}},\ \bibinfo
  {pages} {022119} (\bibinfo {year} {2013})}\BibitemShut {NoStop}%
\end{thebibliography}%

\end{document}